%% file: main.tex
\documentclass[prx,reprint,twocolumn,showpacs,superscriptaddress]{revtex4-1}
\usepackage{amsmath,amssymb,bm,mathrsfs,graphicx, braket,amsthm,enumerate}
\usepackage[colorlinks=true,citecolor=blue,linkcolor=blue]{hyperref}
\usepackage{longtable}
\usepackage{multirow}
\usepackage[all,cmtip]{xy}
\usepackage[normalem]{ulem}
\usepackage[usenames,dvipsnames]{color}

\renewcommand{\vec}[1] {{\rm{\bf #1}}}
\newcommand{\vechat}[1]{\hat{\bm{#1}}}

\begin{document}
\title{Symmetry-based Indicators of Band Topology in the 230 Space Groups}
\begin{abstract}
The interplay between symmetry and topology leads to a rich variety of electronic topological phases, protecting states such as the topological insulators and Dirac semimetals. Previous results, like the Fu-Kane parity criterion for inversion-symmetric topological insulators, demonstrate that symmetry labels can sometimes unambiguously indicate underlying band topology. Here we develop a systematic approach to expose all such symmetry-based indicators of band topology in all the 230 space groups. This is achieved by first developing an efficient way to represent band structures in terms of elementary basis states, and then isolating the topological ones by removing the subset of atomic insulators, defined by the existence of localized symmetric Wannier functions.  Aside from encompassing all earlier results on such indicators, including in particular the notion of filling-enforced quantum band insulators, our theory identifies symmetry settings with previously hidden forms of band topology, and can be applied to the search for topological materials.
\end{abstract}

\author{Hoi Chun Po}
\author{Ashvin Vishwanath}\thanks{Corresponding author: avishwanath@g.harvard.edu}
\affiliation{Department of Physics, University of California, Berkeley, CA 94720, USA}
\affiliation{Department of Physics, Harvard University, Cambridge MA 02138}

\author{Haruki Watanabe} 
\affiliation{Department of Applied Physics, University of Tokyo, Tokyo 113-8656, Japan}

\maketitle

\section{Introduction}
The discovery of topological insulators (TIs) has reinvigorated the well established theory of electronic band structures \cite{TI2013,BernevigBook}.
Exploration along this new dimension has led to an ever-growing arsenal of topological materials, which include, for instance, topological (crystalline) insulators \cite{RMP_TI,FuPRLTCI, TCI_SnTe}, quantum anomalous Hall insulators \cite{QAH}, and Weyl and Dirac semimetals \cite{RevModPhys.88.021004}. 
Such materials possess unprecedented physical properties, like quantized response and gapless surface states, that are robust against all symmetry-preserving perturbations as long as a band picture remains valid \cite{TI2013, BernevigBook, RevModPhys.88.021004}.
 
Soon after these new developments, it was realized that symmetries of energy bands, a thoroughly studied aspect of band theory, is also profoundly intertwined with topology. This is exemplified by the celebrated Fu-Kane criterion for inversion-symmetric materials, which demarcates TIs from trivial insulators using only their parity eigenvalues \cite{Fu-Kane}. 
This criterion, when applicable, greatly simplifies the topological analysis of real materials, and underpins the theoretical prediction and subsequent experimental verification of many TIs \cite{HgTe_Th, HgTe_Exp,Fu-Kane, BiSb_Exp, BiSe_Exp}.

It is of fundamental interest to obtain results akin to the Fu-Kane criterion in other symmetry settings. Early generalizations in systems with broken time-reversal (TR) symmetry in 2D constrained the Chern number ($C$). The eigenvalues of an $n$-fold rotation was found to determine $C$ modulo $n$ \cite{Ari, Bernevig, Chern_Rotation}.  This is characteristic of a symmetry-based indicator of topology--when the indicator is nonvanishing, band topology is guaranteed, but certain topological phases (i.e $C$ a multiple of $n$ in this context) may be invisible to the indicator.  In 3D, it was also recognized that spatial inversion alone can protect nontrivial phases. A new feature here is that these phases do not host protected surface states, since inversion symmetry is broken at the surface, but they do represent distinct phases of matter. For example they possess nontrivial Berry phase structure in the Brillouin Zone which leads to robust entanglement signatures \cite{Ari10, Ari, Bernevig, WannierSheet} and, in some cases, quantized responses \cite{Ari, Bernevig, YuanMing}. Interestingly, in the absence of TR invariance the inversion eigenvalues (i.e., parities) can also protect Weyl semimetals \cite{Ari, Bernevig}, which informed early work on materials candidates \cite{PhysRevB.83.205101}. Hence, these symmetry-based indicators are relevant both to the search for nontrivial insulating phases, and also to the study of topological semimetals. It is also important to note that the goal here is distinct from the classification of topological phases, but is instead to identify signatures of band topology in the symmetry transformations of the state.

An important open problem is to extend these powerful symmetry indicators for band topology to all space groups (SGs). Earlier studies have emphasized the topological perspective, which typically rely on constructions that are specifically tailored to particular band topology of interest \cite{Fu-Kane, Chern_Rotation, Ken2015, Ken2016}. {While some general mathematical frameworks have been developed \cite{FreedMoore, Gomi, Ken2017}, obtaining a full list of concrete results from such an approach} faces an inherent challenge stemming from the sheer multitude of physically relevant symmetry settings--there are 230 SGs in 3D, and each of them is further enriched by the presence or absence of both spin-orbit coupling and TR symmetry. 

A complementary, symmetry-focused perspective leverages the existing exhaustive results on band symmetries \cite{Bradley, Bilbao} to simplify the analysis. Previous work along these lines has covered restricted cases \cite{Ari, Bernevig, Robert-JanSlager, Combinatorics}. For instance, in Ref.~\cite{Combinatorics}, which focuses on systems in the wallpaper groups without any additional symmetry, such an approach was adopted to help develop a more physical understanding of the mathematical treatment of Ref.~\cite{FreedMoore}.
However, the notion of nontriviality is a relative concept in these approaches. While such formulation is well-suited for the study of phase transitions between different systems in the same symmetry setting, it does not always indicate the presence of underlying band topology.
As an extreme example, such classifications generally regard atomic insulators with different electron fillings as distinct phases, although all the underlying band structures are topologically trivial. 

Here, we adopt a symmetry-based approach that focuses on probing the underlying band topology. At the crux of our analysis is the observation that topological band structures arise whenever there is a mismatch between momentum-space and real-space solutions to symmetry constraints \cite{Z2Wannier, SA}. To quantitatively expose such mismatches, we first develop a mathematical framework to efficiently analyze all possible band structures consistent with any symmetry setting, and then discuss how to identify the subset of band structures arising from atomic insulators, which are formed by localizing electrons to definite orbitals in real space. The mentioned mismatch then follows naturally as the quotient between the allowed band structures and those arising from real-space specification. We compute this quotient for all 230 SGs with or without spin-orbit coupling and/ or time-reversal symmetry. Using these results, we highlight symmetry settings suitable for finding topological materials, including both insulators and semimetals. In particular, we will point out that, in the presence of inversion symmetry, stacking two strong 3D TIs will not simply result in a trivial phase, despite all the $\mathbb Z_2$ indices have been trivialized. Instead, it is shown to produce a quantum band insulator \cite{SA} which can be diagnosed through its robust gapless entanglement spectrum. 

\section{Overview of strategy}
Our major goal is to systematically quantify the mismatch between momentum-space and real-space solutions to symmetry constraints in free-electron problems \cite{SA}.
While atomic insulators, which by definition possess localized symmetric Wannier orbitals, can be understood from a real-space picture with electrons occupying definite positions as if they were classical particles, topological band structures (that are intrinsic to dimensions greater than one) do not admit such a description. 
Whenever there is an obstruction to such a real-space reinterpretation, despite the presence of a band gap, the insulating state can only be described  through the quantum interference of electrons, and we refer in general to such systems as quantum band insulators (QBIs). While all topological phases such as Chern insulators, weak and strong $\mathbb{Z}_2$ topological insulators and topological crystalline insulators with protected surface states in $d>1$ are QBIs, more generally, QBIs may or may not have surface states, as the protecting symmetries are not necessarily compatible with any surface termination. Nonetheless, they represent distinct phases of matter, and showcase nontrivial Berry phase in the Brillouin Zone \cite{BerryPhase_TCI}, robust entanglement signatures \cite{Ari10, Ari, Bernevig, SA}, and sometimes quantized responses \cite{Ari, Bernevig, YuanMing}.

Building on this insight, we develop an efficient strategy for identifying topological materials indicated by symmetries. We will first outline a simple framework to organize the set of all possible band structures using only their symmetry labels. 
By extending the ideas in Refs.~\cite{Ari, Combinatorics} and allowing for both addition (stacking) as well as formal subtractions of bands, we show that band structures can be conveniently represented in terms of a special type of Abelian group, which is simply called a `lattice' in mathematical nomenclature.
Next, we systematically isolate topological band structures by quotienting out those that can arise from a Wannier description. We perform this computation for all of the 230 SGs, covering all cases with or without TR symmetry and spin-orbit coupling. Our scheme automatically encompasses all previous results concerning symmetry indicators of band topology, including in particular the Fu-Kane criterion, the relation between Chern numbers and rotation eigenvalues, and the inversion-protected nontrivial phases.

As such band topology is uncovered from the symmetry representations of the bands, we will refer to it as being represented-enforced. We also discuss a more constrained approach where one first specifies the microscopic lattice degrees of freedom. This is relevant to materials where a hierarchy of energy scales isolates a group of atomic orbitals. We find examples where these constraints lead to semimetallic behavior despite band insulators at the same filling are symmetry-allowed. We will refer to these as lattice-enforced semimetals and give a concrete tight-binding example of them. Generalizations of these approaches should aid in the discovery of experimentally relevant topological semimetals and insulators.

Finally, we make two remarks. First, we ignore electron-electron interactions. Second, while our approach is applicable in any dimension, in the special case of 1D even topological phases are smoothly connected to atomic insulators \cite{NRead}, and therefore are regarded as trivial within our framework. These states, and their descendants in higher dimensions, are collectively known as frozen-polarization insulators \cite{Ari}, and will be absent from our discussion.  

\section{Band structures form an abelian group
\label{sec:BS_Zd}}
\begin{figure}[h]
\begin{center}
{\includegraphics[width=0.48 \textwidth]{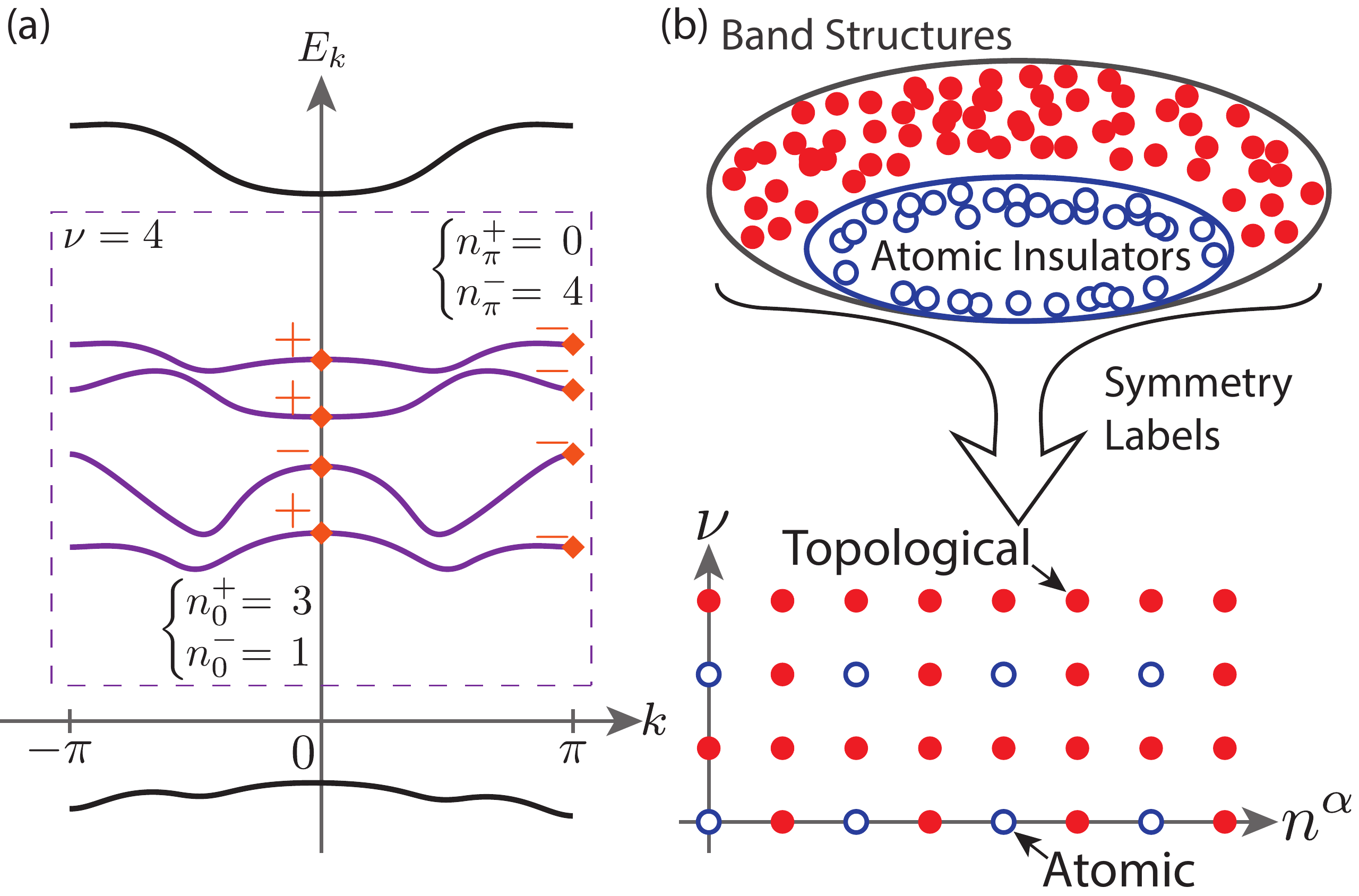}} 
\caption{
{{\bf Symmetry-based indicators of band topology.}
(a) Symmetry labelling of bands in a 1D inversion-symmetric example. $k_0=0,~\pi$ are high-symmetry momenta, where the bands are either even ($+$) or odd ($-$) under inversion symmetry (orange diamonds). 
From a symmetry perspective, a target set of bands (purple and boxed)  separated from all others by band gaps can be labeled by the multiplicities of the two possible symmetry representations, which we denote by the integers $n^{\pm}_{k_0} $. Note that such labelling is insensitive to the detailed energetics within the set. In addition, the set is also characterized by the number of bands involved, which we denote by $\nu$. Altogether, the set is characterized by five integers, but are not independent as they are subjected to the constraints $\nu = n^+_0 + n^-_0 = n^+_\pi + n^-_\pi$.
(b) Symmetry labels like those described in (a) can be similarly defined for systems symmetric under any of the 230 space groups in three dimensions. Using such labels, one can reinterpret the set of band structures as an abelian group. 
This is schematically demonstrated through the two labels $\nu$ and $n^\alpha$, which organize the set of all possible band structures into a two-dimensional lattice. Note that the dimension of this lattice is given by the number of independent symmetry labels, and is a property of the symmetry setting at hand. Organized this way, the band structures corresponding to atomic insulators, which are trivial by our definition, will generally occupy a sublattice. Any band structure that does not fall within this sublattice necessarily possesses nontrivial band topology.}
\label{fig:Scheme}
 }
\end{center}
\end{figure}

We will first argue that the possible set of band structures symmetric under an SG $\mathcal G$ can be naturally identified as the group $\mathbb Z^{d_{\rm BS}}\equiv \mathbb Z \times \mathbb Z\times\dots\times \mathbb Z$, where $d_{\rm BS}$ is a positive integer that depends on both $\mathcal G$ and the spin of the particles
(Fig.~\ref{fig:Scheme}). We will first set aside TR symmetry, and later discuss how it can be easily incorporated into the same framework. The discussion in this section follows immediately from well established results concerning band symmetries \cite{Bradley}, and the same set of results was recently utilized in Ref.~\cite{Combinatorics} to discuss an alternative way to understand the more formal classification in Ref.~\cite{FreedMoore}. Although there is some overlap between the discussion here and that in Ref.~\cite{Combinatorics}, we will focus on a different aspect of the narration: Instead of being solely concerned with the values of $d_{\rm BS}$, we will be more concerned with utilizing this framework to extract other physical information about the systems. As an intermediate step we also perform the first computation of $d_{\rm BS}$ for the 230 space groups.  The results for TR invariant systems are summarized in Tables~\ref{tab:Spinful_TRd}, and those without TR symmetry are tabulated in Appendix \ref{app:SuppTab}.

We begin by reviewing some basic notions using a simple example. Consider free electrons in a 1D, inversion-symmetric crystal. The energy bands $E_m(k)$ are naturally labeled by the band index $m$ and the crystal momentum $k \in (-\pi,\pi]$. Since inversion $P$ flips $k\leftrightarrow -k$, the Bloch Hamiltonian $H(k)$ is symmetric under $P H(k) P^{-1} = H(-k)$, which implies $E_m(k) = E_m(-k)$, and the wavefunctions are similarly related. 
The two momenta $k_0=0$ and $\pi$ are special as they satisfy $P (k_0) = k_0$ (up to a reciprocal lattice vector). As such, the symmetry constraint imposed by $P$ becomes a local constraint at $k_0$, which implies the wavefunctions $\psi_{m}(k_0)$ (generically) furnish irreducible representations (irreps) of $P$: $\psi_m^\dagger (k_0) P \psi_m(k_0) = \zeta_{m}(k_0)$ with  $\zeta_m (k_0) = \pm 1$.

The parities $\zeta_m(k_0)= \pm 1$ can be regarded as local (in momentum space) symmetry labels for the energy band $E_m(k)$, and such labels can be readily lifted to a global one assigned to any set of bands separated from others by a band gap.
We will refer to such sets of bands as `band structures' (BSs), though, as we will explain, caution has to be taken when this notion is used in higher dimensions.
Insofar as symmetries are concerned, we can label the BS by its filling, $\nu$, together with the four non-negative integers, $n^{\pm }_{k_0}$, corresponding to the multiplicity of the irrep $\pm$ at the momenta $k_0 = 0$ or $\pi$ (Fig.~\ref{fig:Scheme}a). Generally, such labels are not independent, since the assumption of a band gap, together with the continuity of the energy bands, cast global symmetry constraints on the symmetry labels. These constraints are known as compatibility relations. For our 1D problem at hand, there are only two of them, which arise from the filling condition: $\nu = n^+_0+ n^-_0 = n^+_{\pi} + n^-_{\pi}$. Consequently the BS is fully specified by three non-negative integers, which we can choose to be $n^+_0$, $n^+_{\pi}$ and $\nu$.

This discussion to this point is similar to that of Ref.~\cite{Combinatorics}, but we now depart from the combinatorics point of view of that work. Instead, similar to Ref.~\cite{Ari} we develop a mathematical framework to efficiently characterize energy bands in terms of their symmetry transformation properties, and  then show that it provides a powerful tool for the analyzing of general band structures.
To begin, we first note that any BS in this 1D, inversion-symmetry problem can be represented by a five-component `vector'  $\vec n  \equiv (n^+_0,n^-_0, n^+_{\pi}, n^-_{\pi},\nu) \in \mathbb Z^{5}_{\geq 0}$, where $\mathbb Z_{\geq 0}$ denotes the set of non-negative integers. 
In addition, $\vec n$ is subjected to the two (independent) compatibility relations.
We can arrange these relations into a system of linear equations and denote them by a $2\times 5$ matrix $\mathcal C$. The admissible BSs then satisfy $\mathcal C \vec n = \vec 0$, and hence $\ker \mathcal C$, the solution space of $\mathcal C$, naturally enters the physical discussion.
For the current problem, $\ker \mathcal C$ is $3$-dimensional, which echoes with the claim that the BS is specified by three non-negative integers.
At this point, however, it is natural to make a mathematical abstraction and lift the physical condition of non-negativity, and thereby define
\begin{equation}\begin{split}\label{eq:BSDef}
\{ {\rm BS} \} \equiv \ker \mathcal C \cap \mathbb Z^D,
\end{split}\end{equation}
where for the 1D problem at hand we have $D=5$.
The main advantage of this abstraction is that, unlike $\mathbb Z^D_{\geq 0}$, $\mathbb Z^D$ is an abelian group, which has simple structures that greatly simplify our forthcoming analysis.
In particular, $\{ {\rm BS} \}$ so defined can be identified with $\mathbb Z^{d_{\rm BS}}$, where $d_{\rm BS}=3$ is the dimension of the solution space $\ker \mathcal C$.
Physically, the addition in $\mathbb Z^{d_{\rm BS}}$ corresponds to the stacking of energy bands.

Next, we generalize the discussion to any SG $\mathcal G$ in three dimensions. (The general framework applies directly to any spatial dimensions; here we focus on three dimensions for physical reasons.) We call a momentum $\vec k$ a high-symmetry momentum if there is any $g\in \mathcal G$ other than the lattice translations such that $g ( \vec k )= \vec k$ (up to a reciprocal lattice vector). We define a `band structure' as `a set of energy bands isolated from all others by band gaps above and below at all high-symmetry momenta'. Note that in 3D, the phrase `all high-symmetry momenta' includes all high-symmetry points, lines and planes.
The discussion for the 1D example carries through, except that one has to consider a much larger zoo of irreps and compatibility relations \cite{Bradley}. 
{
For brevity, we only summarize the key generalizations below, and relegate a detailed discussion to Appendices \ref{app:SymBS} and \ref{app:AI}.

Similar to the 1D example, in the general 3D setting a collection of integers, corresponding to the multiplicities of the irreps in the BS, is assigned to each high-symmetry momentum. By the gap condition, these integers are invariant along high-symmetry lines. In addition, any pair of symmetry-related momenta will share the same labels. Altogether, we see that the symmetry content of a BS, together with the number of bands $\nu$, is similarly specified by a finite number of integers, which forms the group $\mathbb Z^D$ under stacking. 
These integers are again subjected to the compatibility relations, which arise whenever a high-symmetry momentum is continuously connected to another with a lower symmetry. By continuity, the symmetry content of the BS at the lower-symmetry momentum is fully specified by that of the higher-symmetry one, giving rise to linear constraints we denote collectively by the matrix $\mathcal C$.
The group $\{ {\rm BS} \}$ is then defined as in Eq.~\eqref{eq:BSDef}, and again we find} 
\begin{equation}\begin{split}\label{eq:BS_Z}
\{ {\rm BS} \} \equiv \ker \mathcal C \cap \mathbb Z^D \simeq \mathbb Z^{d_{\rm BS}},
\end{split}\end{equation}
where as before $d_{\rm BS} = \dim\ker \mathcal C$. {Note that this result has a simple geometric interpretation: From the definition Eq.~\eqref{eq:BSDef}, we can picture $\ker \mathcal C$ as a $d_{\rm BS}$-dimensional hyperplane slicing through the hypercubic lattice $\mathbb Z^D$ embedded in $\mathbb R^D$ (Appendix \ref{app:X_BS}). This gives rise to the sublattice $\mathbb Z^{d_{\rm BS}}$ (Fig.~\ref{fig:Scheme}b).
}

While Eq.~\eqref{eq:BS_Z} follows readily from definitions, it has interesting physical implications. 
As a group, $\mathbb Z^{d_{\rm BS}}$ is generated by $d_{\rm BS}$ independent generators. In the additive notation, natural for an abelian group, we can write the generators as $\{ \vec b_{i}~:~ i = 1,\dots, d_{\rm BS}\}$, and for any given BS we can expand it similar to elements in a vector space
\begin{equation}\begin{split}\label{eq:BSExp}
{\rm BS} = \sum_{i=1}^{d_{\rm BS}} m_i \vec b_i,
\end{split}\end{equation}
where $m_i \in \mathbb Z$ are uniquely determined once the basis is fixed. 
Therefore, full knowledge of $\{ {\rm BS} \}$ is obtained once the $d_{\rm BS}$ generators $\vec b_i$ are specified.

{Having shown that $\{ {\rm BS}\}$ is a well-defined mathematical entity and identified its general structure, it remains to connect it to the study of physical band structures. Recall that in motivating the definition Eq.~\eqref{eq:BSDef}, we have lifted the physical condition that all irreps must appear a non-negative number of times. This implies any physical band structure must  correspond to elements in the subset $\{ {\rm BS} \}_{\rm P} \equiv \ker C \cap \mathbb Z^D_{\geq 0} \subset \{ {\rm BS}\}$. Nonetheless, any element of $\{ {\rm BS} \}_{\rm P}$ enjoys the properties of any general element of $ \{ {\rm BS}\}$, and in particular the expansion Eq.~\eqref{eq:BSExp} applies. Another important observation is that all elements in $\{ {\rm BS} \}_{\rm P} $ corresponds to some physical band structures. This is shown in Appendix \ref{app:X_BS}, where we argue that the gap condition imposed in the definition of BSs ensures such correspondence.
}

So far, we have not addressed the effect of TR symmetry, which, being anti-unitary, does not lead to new irreps when it is incorporated \cite{Bradley}. Instead, TR symmetry could force certain irreps to become paired with either itself or another (Appendix \ref{app:SymBS}). When under TR an irrep $\alpha$ at $\vec k$ is paired with a different irrep $\beta$ at $\vec k'$, where $\vec k' = \vec k$ or $\vec k' = - \vec k$, we simply add to $\mathcal C$ an additional compatibility relation $n_{\vec k}^{\alpha} = n_{\vec k'}^{\beta}$; when $\alpha$ is paired with itself, we demand $\alpha$ to be an even integer, which can be achieved by redefining ${\tilde n}_{\vec k}^{\alpha} \equiv  n_{\vec k}^{\alpha}/2$ and a corresponding rewriting of $\mathcal C$ in terms of $\tilde {\vec n}$.
{Note also that TR is not included in our definition of high-symmetry momenta, although we will always take Kramers degeneracy in spin-orbit-coupled systems into account.} 
In brief, all TR constraints can be incorporated in a simple manner in the described framework, and therefore we will treat them on equal footings with SG constraints from now on.

\section{Atomic insulators and mismatch classification}
While we have provided a systematic framework to probe the structure of $\{ {\rm BS } \}$, much insight can be gleaned from a study of atomic insulators (AIs). AIs correspond to band insulators constructed by first specifying a symmetric set of lattice points in real space, and then fully occupying a set of local orbitals on each of the lattice sites. 
The possible set of AIs can be easily read off from tabulated data of SGs \cite{ITC, Bilbao} as we explain in (Appendix \ref{app:AI}).
In addition, once the real-space degrees of freedom are specified one can readily compute the corresponding element in $\{ {\rm BS} \}$. 
As stacking two AIs lead to another AI, we see that $\{ {\rm AI} \} \leq \{ {\rm BS} \}$ as groups. Any subgroup of $\mathbb Z^{d_{\rm BS}}$ is again a free, finitely-generated abelian group, and therefore we conclude
\begin{equation}\begin{split}\label{eq:AI_Expand}
\{ {\rm AI} \} \simeq \mathbb Z^{d_{\rm AI}} \equiv \left \{ \sum_{i=1}^{d_{\rm AI}} m_i \vec a_{i}~:~ m_i\in \mathbb Z \right\},
\end{split}\end{equation}
where we denote by $\{ \vec a_i\}$ a complete set of basis for $\{ {\rm AI} \}$. 

Once $\{ {\rm BS} \}$ and $\{ {\rm AI} \}$ are separately computed, it is straightforward to evaluate the quotient group (Appendix \ref{app:X_BS})
\begin{equation}\begin{split}\label{eq:}
X_{\rm BS} \equiv \frac{\{ {\rm BS} \}}{\{ {\rm AI} \}}. 
\end{split}\end{equation}
Physically, an entry in  $X_{\rm BS}$ corresponds to an infinite class of BSs that, while distinct as elements of $\{ {\rm BS} \}$, only differ from each other by the stacking of an AI. By definition, the entire subgroup $\{ {\rm AI} \}$ collapses into the trivial element of $X_{\rm BS}$. Conversely, any nontrivial element of $X_{\rm BS}$ corresponds to BSs that cannot be be interpreted as AIs, and therefore $X_{\rm BS}$ serves as a symmetry indicator of topological BSs.

Following the described recipe, we compute $\{ {\rm AI} \}$, $\{ {\rm BS} \}$, and $X_{\rm BS}$ for all 230 SGs in the four symmetry settings mentioned.  
$X_{\rm BS}$ for spinful fermions with TR symmetry, relevant for real materials with or without spin-orbit coupling and no magnetic order, are tabulated in Tables~\ref{tab:Spinful_TRX} and \ref{tab:Spinless_TRX}.  The results for the non-TR-symmetric settings are presented in Appendix \ref{app:SuppTab}. 
{We also note that the corresponding results for quasi-1D and 2D systems, described respectively by rod and layer group symmetries, can be readily obtained \cite{ITCE, Bilbao}. The results are also presented in Appendix \ref{app:SuppTab}. In particular, we found $X_{\rm BS} = \mathbb Z_1$, the trivial group, for all quasi-1D systems. This is consistent with the picture that topological band structures in 1D can be understood as frozen polarization states, which are AIs and hence trivial in our definition.
}

An interesting observation from this exhaustive computation is the following: for all the symmetry settings considered, we found $d_{\rm BS} = d_{\rm AI}$ (Tables~\ref{tab:Spinful_TRd} and \ref{tab:Spinless_TRd}), and therefore $X_{\rm BS}$ is always a finite abelian group. 
Equivalently, when only symmetry labels are used in the diagnosis, a BS is nontrivial precisely when it can only be understood as a fraction of an AI.
In addition, $d_{\rm BS} = d_{\rm AI}$ implies that a complete set of basis for $\{ {\rm BS}\}$ can be found by studying combinations of AIs, similar to Eq.~\eqref{eq:AI_Expand} but with a generalization of the expansion coefficients $m_i \in \mathbb Z$ to $q_{i} \in \mathbb Q$, subjected to the constraint that the sum remains integer-valued.  Although the full set of compatibility relations is needed in our computation establishing $d_{\rm BS} = d_{\rm AI}$, using our results $\{ {\rm BS} \}$ can be readily computed directly from $\{ {\rm AI} \}$. Since $\{ {\rm BS} \}$ can be easily found this way, we will refrain from providing its complete list.

To illustrate the ideas more concretely, we discuss a simple example concerning spinless fermions without TR invariance, but symmetric under SG {106}.  In this setting, $d_{\rm BS} = d_{\rm AI} = 3$, and $\vec a_1$, one of the three generators of $\{ {\rm AI}\}$, has the property that all irreps appear an even number of times, while the other two generators contain some odd entries.  Now consider $\vec{b}_1\equiv \vec{a}_1/2$, which is still integer-valued. Clearly, the linearity of the the problem implies $\vec b_1$ satisfies all symmetry constrains, and therefore $\vec b_1 \in \{ \rm BS\}$. However, $\vec b_1\not \in \{ {\rm AI}\}  \equiv \{\sum_{i=1}^{3}m_1 \vec a_1 ~:~ m_i \in \mathbb Z\}$, and therefore $\vec b_1$ corresponds to a quantum BS, and indeed it is a representative for the nontrivial element of $X_{\rm BS}=\mathbb{Z}_2$. In addition, if we consider a tight-binding model with representation content corresponding to $\vec a_1$, the decomposition $\vec a_1 = \vec b_1 + \vec b_1$ implies it is possible to open a band gap at all high-symmetry momenta at half filling, and thereby realizing the quantum BS $\vec b_1$. 
It turns out that, in fact, $\vec b_1$ corresponds to a filling-enforced QBI (feQBI) \cite{SA}. We will elaborate further on this point in Appendix \ref{app:feSM}.

Before we move on to concrete applications of our results, we pause to clarify some subtleties in the exposition. Recall that the notion of BS is defined using the presence of band gaps at all high-symmetry momenta. Generally, however, there can be gaplessness in the interior of the Brillouin zone that coexist with our definition of BS. While in some cases such gaplessness is accidental in nature, in the sense that it can be annihilated without affecting the BS, in some more interesting cases it is enforced by the specification of the symmetry content. 
This was pointed out in Refs.~\cite{Ari,Bernevig} for inversion-symmetric systems without TR symmetry, where certain assignment of the parity eigenvalues ensures the presence of Weyl points at some generic momenta. 
When a nontrivial element in $X_{\rm BS}$ can be insulating, we refer to it as as a representation-enforced QBI (reQBI); when it is necessarily gapless, we call it a representation-enforced semimetals (reSM).
We caution that $X_{\rm BS}$ will naturally include both reQBIs and reSMs, although some symmetry settings naturally forbid the notion of reSMs. In fact, one can show that their individual diagnoses are related by $X_{\rm SM} = X_{\rm BS}/ X_{\rm BI}$ (Appendix \ref{app:X_Relate}). Hence, given an entry of $X_{\rm BS}$ one has to further decide whether it corresponds to a reSM or a reQBI. 
{In Appendix \ref{app:X_Relate}, we provide general arguments on the existence of reSMs for systems with significant spin-orbit coupling and TR symmetry.}

In addition, we also note that while every BS belonging to a nontrivial class of $X_{\rm BS}$ is necessarily nontrivial, some systems in the trivial class can also be topological.
By definition, an element of BS in the trivial class of $X_{\rm BS}$ can be written as a stacking of atomic insulators. If the stacking necessarily involves at least one negative integer coefficient, the BS element is still topologically nontrivial. Some of the feQBIs discussed in Ref.~\cite{SA} also fall into this category. 

Alternatively, when the topological nature of the band structure is undetectable using only symmetry labels, say for the tenfold-way phases in the absence of any spatial symmetries, the system belongs to the trivial element of $X_{\rm BS}$ despite it is topological.
{
As an example, consider a two-dimensional system with only lattice translation symmetries. For such systems, the K-theory classification of band insulators in Refs.~\cite{FreedMoore,Combinatorics,Ken2017} gives $\mathbb Z^2$, where the two factors correspond respectively to the electron filling (i.e., number of bands) and the Chern number. 
In contrast, within our approach we find $\{ {\rm BS}\} = \{ {\rm AI}\}= \mathbb Z$, as in this setting the only symmetry label is the filling, which cannot detect the Chern number of the bands. Furthermore, as there exists AI for any filling  $\nu$, we find $X_{\rm BS} = \mathbb Z_1$, the trivial group. 

Finally, we note that all nontrivial entries in $X_{\rm BS}$ have physical representatives, i.e.,~any nontrivial class in $X_{\rm BS}$ can be represented by some physical band structures. The proof of this claim will require a small technical corollary concerning the properties of AIs, and we relegate the detailed discussion to Appendix \ref{app:X_BS}.}

\onecolumngrid

\begin{center}
\begin{table}[h]
\caption{
\bf Characterization of band structures for systems with time-reversal symmetry and significant spin-orbit coupling.
\label{tab:Spinful_TRd}}
\begin{tabular}{c|c} \hline \hline
$d$ & Space groups \\
\hline
$1$  &  1, 3, 4, 5, 6, 7, 8, 9, 16, 17, 18, 19, 20, 21, 22, 23, 24, 25, 26, 27, 28, 29, 30, 31, 32, 33, 34\\
     &  35, 36, 37, 38, 39, 40, 41, 42, 43, 44, 45, 46, 76, 77, 78, 80, 91, 92, 93, 94, 95, 96, 98, 101, 102, 105, 106,\\
     &  109, 110, 144, 145, 151, 152, 153, 154, 169, 170, 171, 172, 178, 179, 180, 181\\
\hline
$2$  &  79, 90, 97, 100, 104, 107, 108, 146, 155, 160, 161, 195, 196, 197, 198, 199, 208, 210, 212, 213, 214\\
\hline
$3$  &  48, 50, 52, 54, 56, 57, 59, 60, 61, 62, 68, 70, 73, 75, 89, 99, 103, 112, 113, 114, 116, 117, 118, 120, 122, 133, 142\\
     &  150, 157, 159, 173, 182, 185, 186, 209, 211\\
\hline
$4$  &  63, 64, 72, 121, 126, 130, 135, 137, 138, 143, 149, 156, 158, 168, 177, 183, 184, 207, 218, 219, 220\\
\hline
$5$  &  11, 13, 14, 15, 49, 51, 53, 55, 58, 66, 67, 74, 81, 82, 86, 88, 111, 115, 119, 134, 136, 141, 167, 217, 228, 230\\
\hline
$6$  &  69, 71, 85, 125, 129, 132, 163, 165, 190, 201, 203, 205, 206, 215, 216, 222\\
\hline
$7$  &  12, 65, 84, 128, 131, 140, 188, 189, 202, 204, 223\\
\hline
$8$  &  124, 127, 148, 166, 193, 200, 224, 226, 227\\
\hline
$9$  &  2, 10, 47, 87, 139, 147, 162, 164, 176, 192, 194\\
\hline
$10$  &  174, 187\\
\hline
$11$  &  225, 229\\
\hline
$13$  &  83, 123\\
\hline
$14$  &  175, 191, 221\\
\hline
\hline
\end{tabular}\\
\begin{flushleft}
$d$: the rank of the abelian group formed by the set of band structures.
\end{flushleft}
\end{table}
\end{center}

\begin{center}
\begin{table}[h]
\caption{
\bf Characterization of band structures for systems with time-reversal symmetry and negligible spin-orbit coupling.
\label{tab:Spinless_TRd}}
\begin{tabular}{c|c} \hline \hline
$d$ & Space groups \\
\hline
$1$  &  1, 4, 7, 9, 19, 29, 33, 76, 78, 144, 145, 169, 170\\
\hline
$2$  &  8, 31, 36, 41, 43, 80, 92, 96, 110, 146, 161, 198\\
\hline
$3$  &  5, 6, 18, 20, 26, 30, 32, 34, 40, 45, 46, 61, 106, 109, 151, 152, 153, 154, 159, 160, 171, 172, 173, 178, 179, 199\\
     &  212, 213\\
\hline
$4$  &  24, 28, 37, 39, 60, 62, 77, 79, 91, 95, 102, 104, 143, 155, 157, 158, 185, 186, 196, 197, 210\\
\hline
$5$  &  3, 14, 17, 27, 42, 44, 52, 56, 57, 94, 98, 100, 101, 108, 114, 122, 150, 156, 182, 214, 220\\
\hline
$6$  &  11, 15, 35, 38, 54, 70, 73, 75, 88, 90, 103, 105, 107, 113, 142, 149, 167, 168, 184, 195, 205, 219\\
\hline
$7$  &  13, 22, 23, 59, 64, 68, 82, 86, 117, 118, 120, 130, 163, 165, 180, 181, 203, 206, 208, 209, 211, 218, 228, 230\\
\hline
$8$  &  21, 58, 63, 81, 85, 97, 116, 133, 135, 137, 148, 183, 190, 201, 217\\
\hline
$9$  &  2, 25, 48, 50, 53, 55, 72, 99, 121, 126, 138, 141, 147, 188, 207, 216, 222\\
\hline
$10$  &  12, 74, 93, 112, 119, 176, 177, 202, 204, 215\\
\hline
$11$  &  66, 84, 128, 136, 166, 227\\
\hline
$12$  &  51, 87, 89, 115, 129, 134, 162, 164, 174, 189, 193, 223, 226\\
\hline
$13$  &  16, 67, 111, 125, 194, 224\\
\hline
$14$  &  49, 140, 192, 200\\
\hline
$15$  &  10, 69, 71, 124, 127, 132, 187\\
\hline
$17$  &  225, 229\\
\hline
$18$  &  65, 83, 131, 139, 175\\
\hline
$22$  &  221\\
\hline
$24$  &  191\\
\hline
$27$  &  47, 123\\
\hline
\hline
\end{tabular}\\
\begin{flushleft}
$d$: the rank of the abelian group formed by the set of band structures.
\end{flushleft}
\end{table}
\end{center}

\begin{center}
\begin{table}[h]
\caption{
\bf Symmetry-based indicators of band topology for systems with time-reversal symmetry and significant spin-orbit coupling.
\label{tab:Spinful_TRX}}
\begin{tabular}{c|c} \hline \hline
$X_{\rm BS}$ & Space groups \\
\hline
$\mathbb Z_{2}$  &  81, 82, 111, 112, 113, 114, 115, 116, 117, 118, 119, 120, 121, 122, 215, 216, 217, 218, 219, 220\\
\hline
$\mathbb Z_{3}$  &  188, 190\\
\hline
$\mathbb Z_{4}$  &  52, 56, 58, 60, 61, 62, 70, 88, 126, 130, 133, 135, 136, 137, 138, 141, 142, 163, 165, 167, 202, 203,\\
                 &  205, 222, 223, 227, 228, 230\\
\hline
$\mathbb Z_{8}$  &  128, 225, 226\\
\hline
$\mathbb Z_{12}$  &  176, 192, 193, 194\\
\hline
$\mathbb Z_{2} \times \mathbb Z_{4}$  &  14, 15, 48, 50, 53, 54, 55, 57, 59, 63, 64, 66, 68, 71, 72, 73, 74, 84, 85, 86, 125, 129,\\
                                      &  131, 132, 134, 147, 148, 162, 164, 166, 200, 201, 204, 206, 224\\
\hline
$\mathbb Z_{2} \times \mathbb Z_{8}$  &  87, 124, 139, 140, 229\\
\hline
$\mathbb Z_{3} \times \mathbb Z_{3}$  &  174, 187, 189\\
\hline
$\mathbb Z_{4} \times \mathbb Z_{8}$  &  127, 221\\
\hline
$\mathbb Z_{6} \times \mathbb Z_{12}$  &  175, 191\\
\hline
$\mathbb Z_{2} \times \mathbb Z_{2} \times \mathbb Z_{4}$  &  11, 12, 13, 49, 51, 65, 67, 69\\
\hline
$\mathbb Z_{2} \times \mathbb Z_{4} \times \mathbb Z_{8}$  &  83, 123\\
\hline
$\mathbb Z_{2} \times \mathbb Z_{2} \times \mathbb Z_{2} \times \mathbb Z_{4}$  &  2, 10, 47\\
\hline
\hline
\end{tabular}\\
\begin{flushleft}
$X_{\rm BS}$: the quotient group between the group of band structures and atomic insulators.
\end{flushleft}
\end{table}
\end{center}

\begin{center}
\begin{table}[h]
\caption{
\bf Symmetry-based indicators of band topology for systems with time-reversal symmetry and negligible spin-orbit coupling.
\label{tab:Spinless_TRX}}
\begin{tabular}{c|c} \hline \hline
$X_{\rm BS}$ & Space groups \\
\hline
$\mathbb Z_{2}$  &  3, 11, 14, 27, 37, 48, 49, 50, 52, 53, 54, 56, 58, 60, 66, 68, 70, 75, 77, 82, 85, 86,\\
                 &  88, 103, 124, 128, 130, 162, 163, 164, 165, 166, 167, 168, 171, 172, 176, 184, 192, 201, 203\\
\hline
$\mathbb Z_{2} \times \mathbb Z_{2}$  &  12, 13, 15, 81, 84, 87\\
\hline
$\mathbb Z_{2} \times \mathbb Z_{4}$  &  147, 148\\
\hline
$\mathbb Z_{2} \times \mathbb Z_{2} \times \mathbb Z_{2}$  &  10, 83, 175\\
\hline
$\mathbb Z_{2} \times \mathbb Z_{2} \times \mathbb Z_{2} \times \mathbb Z_{4}$  &  2\\
\hline
\hline
\end{tabular}\\
\begin{flushleft}
$X_{\rm BS}$: the quotient group between the group of band structures and atomic insulators.
\end{flushleft}
\end{table}
\end{center}

\twocolumngrid

\section{Quantum band insulators in conventional settings}
Having derived the general theory for finding symmetry-based indicators of band topology, we now turn to applications of the results.
As a first application, we utilize the results in Table~\ref{tab:Spinful_TRX} to look for reQBIs that are not diagnosed by previously available topological invariants.
In particular, we will focus on an interesting result concerning one of the most well studied symmetry setting: materials with significant spin-orbit coupling symmetric under TR, lattice translations and inversion (SG {2}).

As shown in Table~\ref{tab:Spinful_TRX}, $X_{\rm BS} = (\mathbb Z_2)^3 \times \mathbb Z_4$ for this setting. Using the Fu-Kane criterion \cite{Fu-Kane}, one can readily verify the strong and weak TIs respectively serve as the generators of the $\mathbb Z_4$ and $\mathbb Z_2$ factors. This identification, however, fails to account for the nontrivial nature of the doubled strong TI, which being a nontrivial element in $\mathbb Z_4$ corresponds to a reQBI. This reQBI has no protected surface states and a trivial magnetoelectric response ($\theta = 0$), and it cannot be directly explained using earlier works focusing on inversion-symmetric insulators \cite{Ari10,Ari,Bernevig}.

\begin{figure}[h]
\begin{center}
{\includegraphics[width=0.48 \textwidth]{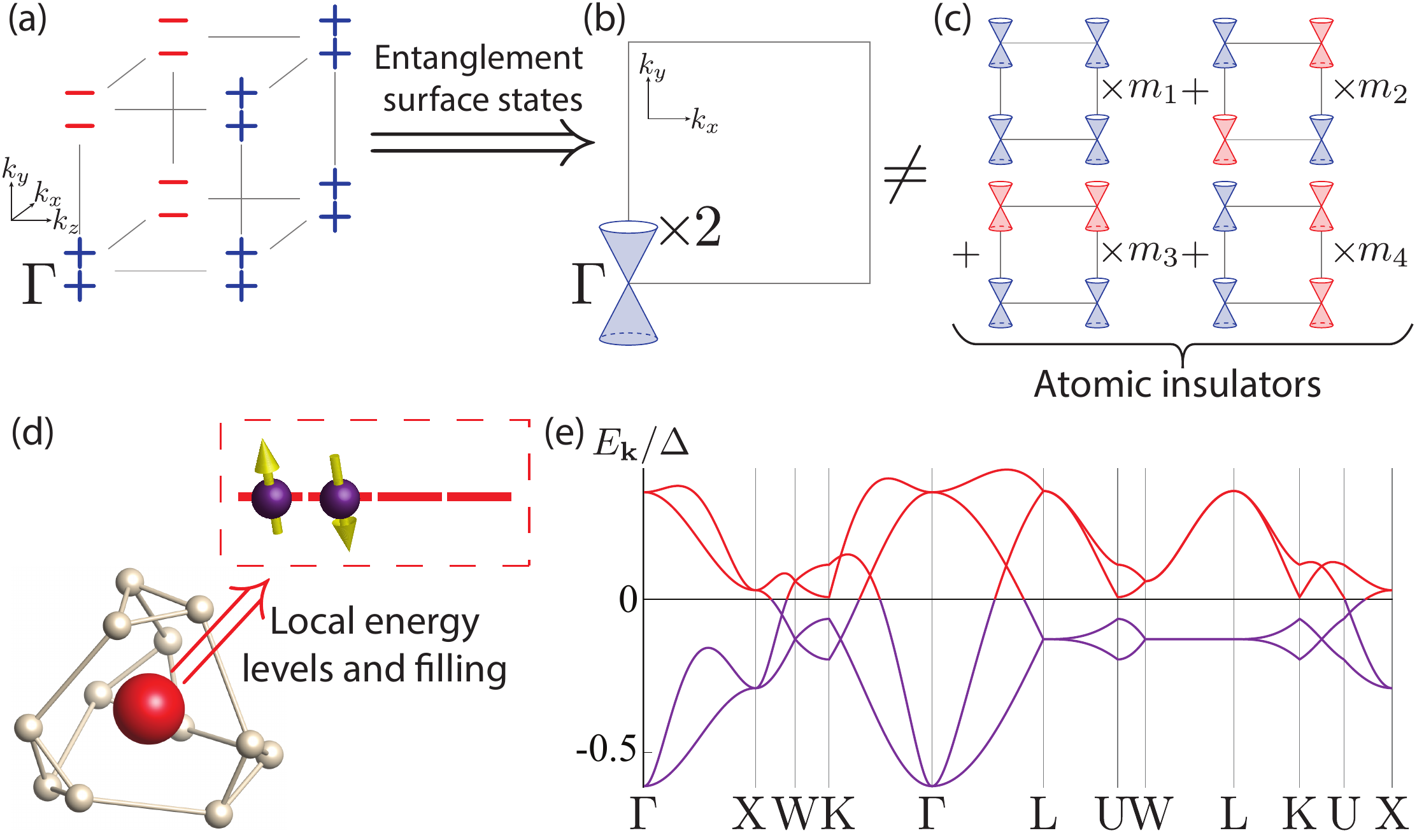}} 
\caption{{
{\bf Examples of topological band structures}
(a-c) A representation-enforced quantum band insulator of spinful electrons with time-reversal and inversion symmetries, dubbed the `doubled strong topological insulator'. (a) Using the Fu-Kane parity criterion \cite{Fu-Kane}, the strong and weak $\mathbb Z_2$ topological insulator (TI) indices can be computed from the the parities of the occupied bands, which we indicate by $\pm$ at the eight time-reversal invariant momenta. 
Shown are the parities of one state from each Kramers pair for a doubled strong TI with four filled bands.
(b) The entanglement spectrum at a spatial cut, parallel to the $x$-$y$ plane and contains an inversion center, features two Dirac cones at $\Gamma $ \cite{Ari10,Ari,Bernevig}. 
Such Dirac cones are known to possess integer-valued charges under the inversion symmetry, and we denote the positively- and negatively-charged cones respectively by blue and red.
(c) Inversion-symmetric atomic insulators feature entanglement surface Dirac cones in general, but their presence depends on the arbitrary choice of the cut.
We find that the possible Dirac-cone arrangement arising from atomic insulators can only be a linear combination of four basic configurations, illustrated as a sum with the integral weights $m_i$. The arrangement in (b) cannot be reconciled with those in (c), confirming the nontriviality of the doubled strong TI.
(d-e) Example of a lattice-enforced semimetal for spinful electrons with time-reversal symmetry.
(d) We consider a site (red sphere) under a local environment (beige) symmetric under the point group $T$, and suppose the relevant local energy levels form the four-dimensional irreducible representation, which is half-filled (boxed).
(e) When the red site sits at the highest-symmetry position of space group 219, the specified local energy levels and filling gives rise to a half-filled eight-band model (each band shown is doubly degenerate). Such (semi-)metallic behavior is dictated by the specification of the microscopic degrees of freedom in this model.
}
\label{fig:Applications}
 }
\end{center}
\end{figure}

Nonetheless, similar to the argument in Refs.~\cite{Ari10,Ari,Bernevig}, the nontrivial nature of the reQBI can be seen from its entanglement spectrum, which exhibits protected gaplessness related to the parity eigenvalues of the filled bands (Fig.~\ref{fig:Applications}a). 
In the present context, we define the entanglement spectrum as the collection of single-particle entanglement energies arising from a spatial cut, which contains an inversion center and is perpendicular to a crystalline axis.
{
Refs.~\cite{Ari10,Ari,Bernevig} showed that the entanglement spectrum of TR and inversion symmetric insulators generally have protected Dirac cones at the TR invariant momenta of the surface Brillouin zone. These Dirac cones carry effective integer charges under inversion symmetry, and as a result they are symmetry-protected. This is observed for the doubled strong TI phase (Fig.~\ref{fig:Applications}b), which has twice the number of Dirac cones as the regular strong TI.
}

Yet, one must use caution in interpreting the nontrivial nature of such entanglement, since inversion-symmetric AIs often have protected entanglement surface states.
They arise whenever the center of mass of an electronic wavefunction is pinned to the entanglement cut, and such states were called frozen-polarization states.
The presence of these entanglement surface modes is dependent on the arbitrary choice of the location of the cut, and therefore are not as robust as the other topological characterizations.
In contrast, since we have already quotient out all AIs in the definition of $X_{\rm BS}$, the reQBI at hand must have a more topological origin. This is verified from the pictorial argument in Figs.~\ref{fig:Applications}a-c, {where we contrast the entanglement spectrum of the doubled strong TI with those that can arise from AIs. Importantly, we see that the total Dirac-cone charge of an AI is always $0 \mod 4$, whereas the doubled strong TI has a charge of $2 \mod 4$. This immediately implies} that the entanglement gaplessness cannot be reconciled with that arising from any AI, {and in fact shows that, in this symmetry setting, the bulk computation of $X_{\rm BS}$ can be reproduced by considering the entanglement spectrum.}
Note that if TR is broken, Kramers paring will be lifted and the irrep content of this reQBI becomes achievable with an AI. This suggests that the reQBI at hand is protected by the combination of TR and inversion symmetry, and in particular it can be interpreted as a TR-symmetric BS mimicking a TR-broken AI through momentum space quantum interference. It is an interesting open question to study whether or not this reQBI has any associated quantized physical response \cite{Ari}. 

Since the strong TI is compatible with any additional spatial symmetry, the argument above is applicable to any centrosymmetric SGs. Indeed, as can be seen from Table~\ref{tab:Spinful_TRX}, all of them have $|X_{\rm BS}| \geq 4$, consistent with our claim. From the same table, we can also observe that, very often, a general weak TI becomes incompatible with a higher degree of spatial symmetries. 
This is related to the number of factors in $X_{\rm BS}$, i.e.~the number of independent generators $N_t$. 
For any centrosymmetric SG, the Fu-Kane criterion dictates that all strong and weak TIs are diagnosable using symmetry labels.
As one such factor is reserved for the strong TI, the SG is compatible with at most $N_t - 1$ independent weak TI phases.
While this has been pointed out for certain cases in the literature \cite{WeakTIConst}, our approach automatically encapsulates some of these result in a simple manner.
Finally, we remark that while the same $X_{\rm BS}$ is found for SG {2} in all the other symmetry settings, their physical interpretations are very different. In particular, the generators of $\mathbb Z_4 < X_{\rm BS}$ corresponds to a reSM in the other settings.

\section{Lattice-enforced semimetals}
As another application of our results, we demonstrate how the structure of $\{ {\rm BS}\}$ exposes constraints on the possible phases of a system arising from the specification of the microscopic degrees.
We will in particular focus on the study of semimetals, but a similar analysis can be performed in the study of, say, reQBIs.

As a warm-up, recall the physics of (spinless) graphene, where specifying the honeycomb lattice dictates that the irrep at the K point is necessarily two-dimensional, and therefore the system is guaranteed to be gapless at half filling. 
Using the structure of $\{ {\rm BS} \}$ we described, this line of reasoning can be efficiently generalized to an arbitrary symmetry setting: Any specification of the lattice degrees of freedom corresponds to an element $\vec A \in \{ {\rm AI}\}$, and one simply asks if it is possible to write $\vec A = \vec B_v + \vec B_c$, where $\vec B_{v,c} \in \{ {\rm BS} \}$ satisfies the physical non-negative condition, such that $\vec B_v$ corresponds to a BS with a specified filling $\nu$. Whenever the answer is no, the system is guaranteed to be (semi-)metallic.
We refer to any such system as a lattice-enforced semimetal (leSM). 
Note that a stronger form of symmetry-enforced gaplessness can originate simply from the electron filling, and such systems were dubbed as filling-enforced semimetal (feSMs) \cite{PNAS,PRL}. We will exclude feSMs from the definition of leSMs, i.e.~we only call a system a leSM if the filling $\nu$ is compatible with some band insulators in the same symmetry setting, but is nonetheless gapless because of the additional lattice constraints.

A preliminary analysis reveals that leSMs abound, especially for spinless systems with TR symmetry.
This is in fact anticipated from the earlier discussions in Refs.~\cite{Zak1999,Zak2000,Zak2001}.
Instead, we will turn our attention to TR-invariant systems with significant spin-orbit coupling, which lies beyond the scope of these earlier studies and oftentimes leads to interesting new physics \cite{TI2013,RevModPhys.88.021004, SA}.
A systematic survey of them will be the focus of another study. 
{Here, we present a proof-of-concept leSM example we found, which arises in TR-symmetric systems in SG 219 ($F\bar{4} 3 c$) with significant spin-orbit coupling.
We will only sketch the key features of the model, and the interested readers are referred to Appendix \ref{app:leSM} for details of the analysis.

We consider a lattice with 2 sites in each primitive unit cell, and that each site has a local environment corresponding to the cubic point group $T$ (Fig.~\ref{fig:Applications}d). We suppose the relevant on-site degrees of freedom transform under the 4D irreducible co-representation of $T$ under TR symmetry \cite{Bradley}, and that the system is at half filling, i.e., the filling is $\nu=4$ electrons per primitive unit cell.}
Although the local orbitals are partially filled, generically a band gap becomes permissible once electron hopping is incorporated. 
Naively, for the present problem this may appear to be the likely scenario, since the momentum-space irreps all have dimensions $\leq 4$ \cite{Bradley} and band insulators are known to be possible at this filling \cite{PNAS,PRL}.
However, using our framework one can prove that no BS is possible for this system at $\nu=4$, which implies there will be irremovable lattice-enforced gaplessness at some high-symmetry line. This is indeed verified in Fig.~\ref{fig:Applications}e, where we plot the band structure obtained from an example tight-binding model (Appendix \ref{app:leSM}).

\section{Discussion}
In this work, we present a simple mathematical framework for efficiently analyzing band structures as entities defined globally over the Brillouin zone. We further utilize this result to systematically quantify the mismatch between the momentum- and real-space descriptions of free electron phases, obtaining a plethora of symmetry settings for which topological materials are possible.

Our results concern a fundamental aspect of the ubiquitous band theory. For electronic problems, we demonstrated the power of our approach by discussing three particular applications, predicting quantum band insulators and semimetals in both conventional and unconventional symmetry settings.
We highlight four interesting future directions below: first, to incorporate the tenfold-way classification into our symmetry-based diagnosis of topological materials \cite{Combinatorics}; second, to discover quantized physical responses unique to the phases we predicted \cite{Ari, Bernevig,YuanMing}; third, to extend the results to magnetic space groups by including the extra compatibility relations \cite{Bradley}; and lastly, to screen materials database for topological materials relying on fast diagnosis invoking only symmetry labels \cite{MaterialsDiscovery}. More broadly, we expect our analysis to shed light on any other fields of studies, most notably photonics and phononics, where the interplay between topology, symmetry and band structures is of interest.

{Note added: Recently, Ref.~\cite{TopoChem} appeared, which has some overlap with the present work, in that it also identifies topological band insulators by contrasting them with atomic insulators. However, the present work differs from Ref.~\cite{TopoChem} in important ways in the formulation of the problem  and the mathematical approach adopted.}

\begin{acknowledgments}
We thank C.-M.~Jian, A.~Turner and M.~Zaletel for insightful discussions and collaborations on earlier works. We also thank C.~Fang for useful discussions.  
AV and HCP were supported by NSF DMR-1411343. AV acknowledges support from a Simons Investigator Award.
H.W. acknowledges support from JSPS KAKENHI Grant Number JP17K17678.
\end{acknowledgments}

\appendix
{\section{Glossary of abbreviations}}
For brevity, we have introduced several abbreviations in the text. For the readers' convenience, we provide a glossary of the less-standard ones here. 

`AI' (atomic insulator): band insulators possessing localized symmetric Wannier functions.

`BS' (band structure): a set of energy bands separated from all others by band gaps above and below at all high-symmetry momenta.

`fe' (filling-enforced): referring to attributes that follow from the electron filling of the system.

`le' (lattice-enforced): referring to attributes that follow from the specification of the microscopic degrees of freedom in the lattice.

`QBI' (quantum band insulators): band insulators, with or without protected surface states, that do not admit any atomic limit provided the protecting symmetries are preserved.

`re' (representation-enforced): referring to attributes that follow from knowledge on the symmetry representations of the energy bands.

`SG' (space group): any one of the 230 spatial symmetry groups of crystals in three dimensions.

`SM' (semimetals): filled bands with gap closings that are stable to infinitesimal perturbations.

`TI' (topological insulator): note that we use this phrase in a restricted sense here, always referring to the $\mathbb Z_2$ topological insulators in two or three dimensions for spin-orbit-coupled system with time-reversal symmetry.\\

\section{Review of Symmetries in Band Structures
\label{app:SymBS}
}
Here, we briefly review some notions and results concerning the consequences of symmetries on band structure. The discussion here will closely mirror that of the 1D example given in the main text, but we will assume a general 3D setting from the outset.

\subsection{Little group and its representation}
Consider a system of noninteracting fermions symmetric under a purely spatial symmetry group $\mathcal G$ with the lattice translation subgroup $T$.  
An element $g\in\mathcal{G}$ that maps a point $\vec{x}\in\mathbb{R}^3$ to $g(\vec{x})=p_g\vec{x}+\vec{t}_g\in\mathbb{R}^3$ may be characterized by an orthogonal matrix $p_g$ and a vector $\vec{t}_g$.
Thanks to the lattice translation $T$, it is natural to label the eigenenergy $E_{m}(\vec k)$ by the band index $m$ and the wavevector $\vec k \in {\rm BZ}$.
The notion of energy bands, however, becomes ambiguous whenever they cross, since the underlying wave function of a single band will generally become discontinuous.
Instead, it is more natural to consider a set of entangled bands, separated from all others by band gaps above and below, as a single entity.

Whether a given set of energy bands can be isolated from the others dictate whether or not the system can be insulating, and therefore is a fundamental question in the study of electronic band structures. 
High-symmetry momenta play a crucial role in this analysis.
The subgroup of $\mathcal G$ that leaves a momentum $\vec k \in {\rm BZ}$ invariant (up to a reciprocal lattice vector $\vec G$) is known as the little group of $\vec k$, which is commonly denoted by $\mathcal G_{\vec k}$~\cite{Bradley}. As a set, we have
\begin{equation}
\mathcal G_{\vec k}=\{g\in \mathcal{G}~:~p_g\vec{k}=\vec{k}+{}^\exists\vec{G}\}.
\end{equation}
By definition, lattice translations $T$ automatically form a subgroup of $\mathcal G_{\vec k}$, and we say $\vec k$ is a high-symmetry momentum whenever $\mathcal G_{\vec k}$ contains any element aside from lattice translations.  Note that when we say `all high-symmetry momenta', we include all high-symmetry points, lines and planes in the BZ.

Generally, a spatial symmetry $g$ demands $E_{m} (p_g \vec k) = E_{m} (\vec k)$, and the corresponding wave functions are similarly related. When $g\in \mathcal G_{\vec k}$, the constraint becomes a local condition in momentum space. This implies the wave functions furnish a representation of $\mathcal G_{\vec k}$~\cite{Bradley}. Such representations encode the transformation properties of the BS at $\vec k$, and can be generally decomposed into a direct sum of irreducible representations (irreps) $u_{\vec k}^\alpha$, which have been exhaustively tabulated in Ref.~\cite{Bradley} for both spinless and spinful fermions. Insofar as symmetry properties are concerned, we can label a BS at $\vec k$ by the set of non-negative integers 
\begin{equation}
\{ n_{\vec k}^{\alpha}~:~\alpha=1,\dots, D_{\vec k} \},
\end{equation}
where $n_{\vec k}^{\alpha} $ denotes the number of times the $\alpha$-th irrep appears in the BS, and $ D_{\vec k}$ denotes the number of irreps. Note that such symmetry labeling makes no reference to the detailed energetics of the system, i.e.~the labels are insensitive to the energetic arrangement of the individual bands within the BS.

\subsection{Types of momenta and compatibility relations}
A priori, a BS will carry independent symmetry labels $\{ n_{\vec k}^{\alpha} \}$ and $\{ n_{\vec k'}^{\beta} \}$ at distinct high-symmetry momenta $\vec k$ and $\vec k'$. However, symmetries and continuity can cast extra constraints on such assignment~\cite{Bradley}.
To see this concretely, let us classify all points in BZ into a finite number of `types': we say $\vec{k}_1$ and $\vec{k}_2$ are of the same type if either $\vec{k}_1$ and $\vec{k}_2$ are symmetry-related, i.e., $\vec{k}_2=p_{g_0}\vec{k}_1+{}^\exists\vec{G}$, or each point of the line connecting $\vec{k}_1$ and $\vec{k}_2$ has the same $\mathcal{G}_{\vec{k}}$ as $\mathcal{G}_{\vec{k}_1}=\mathcal{G}_{\vec{k}_2}$.
For example, imagine a line invariant under a $n$-fold rotation (or screw) $C_n$. Any point on this line (except for possibly the end points) has the identical little group generated by $C_n$ and hence is of the same type. 
Similar situation arises for a plane invariant under a mirror (or a glide). 
By assumption, the existence of band gaps imply the representation content of a BS is invariant within such high-symmetry lines or planes, and therefore it suffices to specify the representations at one arbitrarily chosen $\vec{k}$ point for each type of momenta in the BZ. 
Therefore, a full set of symmetry labels of a BS can be obtained by specifying only those arising from a finite number of representatives. 
In addition, the BS is also labeled by the total number of bands in the BS, which we will denote by $\nu$. Note that physically $\nu$ is simply the electron filling of the system when BS is interpreted as the set of filled bands in the full system.
Aggregating all the labels  $\{ n_{\vec k}^{\alpha} \}$ and $\nu$ into a single quantity, we denote the representation content of a BS by $\vec n$, a set of $D$ non-negative integers where $D =1+\sum_{\vec k}D_{\vec k}$. Here, we have chosen one arbitrary $\vec{k}$ for each type of the $\vec{k}$ vectors in the BZ.

Next, we note that a general $\vec n$ may not be compatible with the continuity of the bands and the gap condition we imposed on BS.
To find the set of admissible labels $\{\vec n\}$, we introduce the notion of compatibility relations.~\cite{Bradley} For any pair of infinitesimally close momenta $\vec k$ and $\vec k + \delta \vec k$, their little groups satisfy $\mathcal G_{\vec k+ \delta \vec k} \leq \mathcal G_{\vec k}$ (assuming a proper choice of labeling for the two momenta). 
Since the irreps at the higher-symmetry momentum $\vec k$ can be decomposed into those of the lower-symmetry ones at $\vec k + \delta \vec k$, the symmetry labels $\{ n^{\alpha}_{\vec k + \delta \vec k} \}$ are fully constrained by $\{ n^{\beta}_{\vec k} \}$. This is captured by the compatibility relations, one for each value of $\alpha$,
\begin{equation}\begin{split}\label{eq:CRk}
n^{\alpha}_{\vec k + \delta \vec k} =  \sum_{\beta} c_{\alpha \beta}^{\vec k,\delta \vec k} n^{\beta}_{\vec k},
\end{split}\end{equation}
where the coefficients $c_{\alpha \beta}^{\vec k,\delta \vec k} $ are non-negative integers. There is also a similar compatibility on the filling: $\nu = \sum_{\alpha} \dim[u^{\alpha}_{\vec k}]\, n^{\alpha}_{\vec k}  $, where $\dim[u^{\alpha}_{\vec k}]$ denotes the dimension of the irrep $\alpha$ (i.e.~the number of bands involved).

As our target is to study BSs as  global entities, it is instructive to collect all compatibility relations into a system of linear equations:
\begin{equation}\begin{split}\label{eq:CR}
\mathcal C \vec n = 0,
\end{split}\end{equation}
where $\mathcal C$ is an integer-valued matrix with coefficients determined by those in Eq.~\eqref{eq:CRk}.
By definition, any BS can be identified with an $\vec n$ satisfying Eq.~\eqref{eq:CR}. Conversely, any set of $D$ non-negative integers $\vec n \in \mathbb Z^{D}_{\geq 0}$ satisfying Eq.~\eqref{eq:CR} can be identified with a physical band structure (\ref{app:X_BS}).

\subsection{Construction of irreps of $\mathcal{G}$}
Given an irrep $u^\alpha_{\vec{k}}$ of $\mathcal{G}_{\vec{k}}$, one can construct an irrep of $\mathcal{G}$, which s called the induced representation.
It is known that every irrep of $\mathcal{G}$ can be constructed in this way~\cite{Bradley}.  Here we review the construction in detail for the case where $u^\alpha_{\vec{k}}$ is a projective representation of $\mathcal{G}_{\vec{k}}$ due to the spin degrees of freedom.

Let $\{|\phi_{i,\vec{k}}^r\rangle\}_{i=1}^{\text{dim}[u_{\vec{k}}^\alpha]}$ be the basis of an irrep $u_{\vec{x}}^r$ of $\mathcal{G}_{\vec{x}}$. Namely, $|\phi_{i,\vec{k}}^r\rangle$ transform under $h, h'\in \mathcal{G}_{\vec{k}}$ as
\begin{eqnarray}
\hat{h}|\phi_{i,\vec{k}}^\alpha\rangle&=&\sum_j|\phi_{j,\vec{k}}^\alpha\rangle [u_{\vec{k}}^\alpha(h)]_{ji},\\
u_{\vec{k}}^\alpha(h)u_{\vec{k}}^\alpha(h')&=&z_{h,h'}u_{\vec{k}}^\alpha(hh'),\\
\hat{h}(\hat{h}'|\phi_{i,\vec{k}}^\alpha\rangle)&=&z_{h,h'}\hat{(hh')}|\phi_{i,\vec{k}}^\alpha\rangle,
\end{eqnarray}
where  $z_{g,g'}=\pm1$ is the projective phase originating from the spin of fermions.  For the spinless case $z_{g,g'}$ should be set to be $1$.

Since elements $g\notin\mathcal{G}_{\vec{k}}$ change $\vec{k}$ to an inequivalent momentum $p_g\vec{k}$ by definition, $\hat{g}|\phi_{i,\vec{k}}^r\rangle$ cannot, in general, be expanded by $\{|\phi_{i,\vec{k}}^r\rangle\}_i$.  The symmetry orbit of $\vec{k}$, $\{\vec{k}_{\sigma}\}_{\sigma=1}^{|\mathcal{G}/\mathcal{G}_{\vec{k}}|}=\{p_g\vec{k}~:~g\in \mathcal{G}\}$, is called the star of $\vec{k}$.  We arbitrarily choose a complete set of representatives $\{g_\sigma\}_{\sigma=1}^{|\mathcal{G}/\mathcal{G}_{\vec{k}}|}$ ($g_{\sigma=1}= e$) of $\mathcal{G}/\mathcal{G}_{\vec{k}}$ that satisfies $\vec{k}_\sigma\equiv p_{g_\sigma}\vec{k}$ and define $|\phi_{i,\vec{k}_\sigma}^\alpha\rangle\equiv\hat{g}_\sigma|\phi_{i,\vec{k}}^\alpha\rangle$.  The set $\{|\phi_{i,\vec{k}_{\sigma}}^r\rangle\}_{i,\sigma}$ serves as the basis of the representation of $\mathcal{G}$.  To see how $|\phi_{i,\vec{k}_\sigma}^\alpha\rangle$ transforms under a general element $g\in\mathcal{G}$, note that $gg_\sigma\in\mathcal{G}$ can be uniquely decomposed into a product of $g_{\sigma'}\in\mathcal{G}/\mathcal{G}_{\vec{k}}$ and $h\in\mathcal{G}_{\vec{k}}$. Therefore, step-by-step, we have
\begin{eqnarray}
&&\hat{g}|\phi_{i,\vec{k}_\sigma}^\alpha\rangle=\hat{g}(\hat{g}_\sigma|\phi_{i,\vec{k}}^\alpha\rangle)\notag\\
&&=z_{g,g_{\sigma}}\hat{(gg_\sigma)}|\phi_{i,\vec{k}}^\alpha\rangle=z_{g,g_{\sigma}}\hat{(g_{\sigma'}h)}|\phi_{i,\vec{k}}^\alpha\rangle\notag\\
&&=\frac{z_{g,g_{\sigma}}}{z_{g_{\sigma'},h}}\sum_{i'}\hat{g}_{\sigma'}|\phi_{i',\vec{k}}^\alpha\rangle [u_{\vec{k}}^\alpha(h)]_{i'i}\notag\\
&&=\frac{z_{g,g_{\sigma}}}{z_{g_{\sigma'},h}}\sum_{i'}\hat{g}_{\sigma'}|\phi_{i',\vec{k}}^\alpha\rangle [u_{\vec{k}}^\alpha(h)]_{i'i}\notag\\
&&=\sum_{\sigma',i'}|\phi_{i',\vec{k}_{\sigma'}}^\alpha\rangle[U^\alpha(g)]_{\sigma'i',\sigma i},
\end{eqnarray}
where
\begin{eqnarray}
[U^\alpha(g)]_{\sigma'j,\sigma i}=\delta_{\vec{k}_{\sigma'},p_g\vec{k}_\sigma}''\frac{z_{g,g_{\sigma}}}{z_{g_{\sigma'},g_{\sigma'}^{-1}gg_\sigma}}[u^\alpha_{\vec{k}}(g_{\sigma'}^{-1}gg_\sigma)]_{ji}
\label{inducedrepofG}
\end{eqnarray}
and $\delta''_{\vec{k}_1,\vec{k}_2}$ is $1$ only when $\vec{k}_1=\vec{k}_2$ modulo a reciprocal lattice vector.  This is the induced representation of $\mathcal{G}$ constructed from $u^\alpha_{\vec{k}}$ of $\mathcal{G}_{\vec{k}}$.  It has a nice property that $U^\alpha$ is irreducible whenever $u^\alpha_{\vec{k}}$ is. 

\subsection{Time-reversal symmetry}
Here we will follow the discussion in Ref.~\cite{Bradley} on the consequence of the TR symmetry $\mathcal{T}$ commuting with every element of $G$.  
We will write $\hat{\mathcal{T}}^2=(\eta_{\mathcal{T}})^{\hat{N}}$ where $\hat{N}$ is the number of fermions and $\eta_{\mathcal{T}}=-1$ ($\eta_{\mathcal{T}}=+1$) for the spinful (spinless) case.  

\subsubsection{General case}
Let us start with a finite group $G$ in general.  Suppose that $\{|i\rangle\}_{i=1}^d$ ($d\equiv\text{dim}[u]$) is a basis of an irrep $u$ of $G$:
\begin{eqnarray}
\hat{g}|i\rangle=|j\rangle u_{ji}(g).
\end{eqnarray}
Then $\{\hat{\mathcal{T}}|i\rangle\}_{i=1}^d$ is a basis of the conjugate representation $u^*(g)$:
\begin{eqnarray}
\hat{g}(\hat{\mathcal{T}}|i\rangle)=(\hat{\mathcal{T}}|j\rangle)u_{ji}^*(g).
\end{eqnarray}
When $u$ and $u^*$ are different irreps, the group $G+\mathcal{T}G$ is simply represented by
\begin{eqnarray}
D(g)=
\begin{pmatrix}
u&0\\
0&u^*
\end{pmatrix},
\quad
D(T)=
\begin{pmatrix}
0&\eta_{\mathcal{T}}\\
1&0
\end{pmatrix}.
\end{eqnarray}
The situation is different when $u$ and $u^*$ are the same irrep, i.e, $u^*(g)=v^\dagger u(g)v$ for a unitary matrix $v$.  Using the definition of $v$ twice, we have $(vv^*)u(g)= u(g)(vv^*)$ for every $g$.  Since $u$ is irreducible, we see $vv^*=\xi\eta_{\mathcal{T}}\openone_d$ for $\xi=\pm1$.  In other words,
\begin{equation}
\label{eq:vT}
v^T=(v^*)^{\dagger}=\xi\eta_{\mathcal{T}} v,\quad \xi, \eta_{\mathcal{T}}=\pm1.
\end{equation}

It is easy to see that the combination $|\bar{i}\rangle\equiv(\hat{\mathcal{T}}|j\rangle)v^\dagger_{ji}$ transforms in the same way as $|i\rangle$ under $G$:
\begin{eqnarray}
\hat{g}|\bar{i}\rangle=|\bar{j}\rangle u_{ji}(g).
\end{eqnarray}
Hence the question is if $|i\rangle$ and $|\bar{i}\rangle$ are the same. To see this, let us evaluate the inner-product $G_{ij}\equiv(|i\rangle, |\bar{j}\rangle)$.
Since $\hat{g}$ is unitary, we have
\begin{eqnarray}
G_{ij}=(|i\rangle, |\bar{j}\rangle)=(\hat{g}|i\rangle, \hat{g}|\bar{j}\rangle)=[u(g)^\dagger G u(g)]_{ij},
\end{eqnarray}
i.e., $G$ and $u(g)$ commute for every $g$ and we must have $G=c\,\openone_d$.  To compute $c$, note that
\begin{eqnarray}
\hat{\mathcal{T}}|i\rangle&=&|\bar{j}\rangle v_{ji},\\
\hat{\mathcal{T}}|\bar{i}\rangle&=&\eta_{\mathcal{T}}|j\rangle (v^T)_{ji}=\xi |j\rangle v_{ji}.
\end{eqnarray}
The second line follows from the first by applying $\hat{\mathcal{T}}$ and using Eq.~\eqref{eq:vT}.  Hence
\begin{eqnarray}
c&=&\frac{1}{d}\sum_{i=1}^d(|i\rangle, |\bar{i}\rangle)=\frac{1}{d}\sum_{i=1}^d(\hat{\mathcal{T}}|\bar{i}\rangle,\hat{\mathcal{T}}|i\rangle)\notag\\
&=&\frac{\xi}{d}(vv^\dagger)_{j,j'}\, (|j'\rangle,|\bar{j}\rangle ) =\xi c.
\end{eqnarray}
Namely, when $\xi=-1$, $c$ vanishes and $\{|i\rangle\}_{i=1}^d$ and $\{\hat{\mathcal{T}}|i\rangle\}_{i=1}^d$ are orthogonal.  On the other hand, when $\xi=1$, $c$ is finite and $\{|i\rangle\}_{i=1}^d$ and $\{\hat{\mathcal{T}}|i\rangle\}_{i=1}^d$ are the same state.

To compute $\xi$ from the character $\chi(g)=\text{tr}[u(g)]$, we use the following identity
\begin{eqnarray}
z_{g,g}u(g^2)&=&u(g)u(g)=[u^*(g)]^*u(g)\notag\\
&=&[v^\dagger u(g)v]^*u(g)=\xi\eta_{\mathcal{T}} v u^*(g)v^*u(g).
\end{eqnarray}
We used Eq.~\eqref{eq:vT} in the last line.  Therefore,
\begin{eqnarray}
\eta_{\mathcal{T}}\frac{1}{|G|}\sum_{g}z_{g,g}\chi(g^2)&=&\xi\frac{1}{|G|}\sum_{g}v_{ij}u_{jk}^*(g)v_{kl}^*u_{li}(g)\notag\\
&&=\xi\frac{1}{d}(vv^\dagger)_{i,i}=\xi.
\end{eqnarray}
where, in the second last line we used the orthogonality of irreps $\sum_{g}u_{ij}^{(\alpha)}(g)u_{kl}^{(\beta)}(g)^*=(|G|/d)\delta_{ik}\delta_{jl}\delta^{\alpha\beta}$.  From this orthogonality, it is also clear that $\sum_{g}z_{g,g}\chi(g^2)=0$ when $u$ and $u^*$ are different irreps.

To summarize, to see if the TR symmetry changes of degeneracy for the irrep $u$ of $G$, one should compute
\begin{eqnarray}
&&\eta_{\mathcal{T}}\frac{1}{|G|}\sum_{g\in G}z_{g,g}\chi(g^2)\notag\\
&&=\begin{cases}
+1& \text{: Degeneracy is unchanged.}\\
-1& \text{: Two $u$'s are paired under TR.}\\
0 & \text{: $u$ and $u^*$ are different and are paired}.
\end{cases}
\end{eqnarray}
This is called Wigner's test in the literature~\cite{Bradley}.  

\subsubsection{TR pairing of $u_{k}^\alpha$}
Let us apply this result to irreps $u_{\vec{k}}^\alpha$ of $\mathcal{G}_{\vec{k}}$.  
If there is no element $g\in \mathcal{G}$ such that $p_g\vec{k}=-\vec{k}$ modulo a reciprocal lattice vector $\vec{G}$, then the TR symmetry just implies an additional degeneracy between $u_{\vec{k}}^\alpha$ and $u^\alpha_{-\vec{k}}$ and this case is easily handled. Thus, suppose that there is at least one $g\in\mathcal{G}$ such that $p_g\vec{k}=-\vec{k}+{}^\exists\vec{G}$.

As explained earlier, an irrep $u_{\vec{k}}^\alpha$ of $\mathcal{G}_{\vec{k}}$ induced an irrep $U^\alpha$ of $\vec{G}$, see Eq.~\eqref{inducedrepofG}. Its character can be expressed as
\begin{eqnarray}
\text{tr}[U^\alpha(g)]&=&\sum_\sigma \delta_{\vec{k}_{\sigma},p_g\vec{k}_\sigma}''\frac{z_{g,g_{\sigma}}}{z_{g_{\sigma},g_{\sigma}^{-1}gg_\sigma}}\text{tr}[u^\alpha_{\vec{k}}(g_{\sigma}^{-1}gg_\sigma)]\notag\\
&=&\sum_\sigma\sum_{h\in\mathcal{G}_{\vec{k}}}\delta_{g,g_{\sigma}hg_{\sigma}^{-1}}\frac{z_{g,g_{\sigma}}}{z_{g_{\sigma},h}}\text{tr}[u^\alpha_{\vec{k}}(h)]\notag\\
&=&\sum_\sigma\sum_{h\in\mathcal{G}_{\vec{k}_\sigma}}\delta_{g,h}\text{tr}[u^\alpha_{\vec{k}_\sigma}(h)]\notag\\
&=&\sum_\sigma\sum_{h\in\mathcal{G}_{\vec{k}_\sigma}}\delta_{g,h}\chi^\alpha_{\vec{k}_\sigma}(h),
\end{eqnarray}
where $u^\alpha_{\vec{k}_\sigma}(h)\equiv\frac{z_{h,g_{\sigma}}}{z_{g_{\sigma},g_\sigma^{-1} g_hg_\sigma}}u^\alpha_{\vec{k}}(g_{\sigma}^{-1}hg_{\sigma})$ and $\chi^\alpha_{\vec{k}_\sigma}(h)=\text{tr}[u^\alpha_{\vec{k}_\sigma}(h)]$.  Hence,
\begin{eqnarray}
&&\frac{1}{|\mathcal{G}|}\sum_{g\in\mathcal{G}}z_{g,g}\text{tr}[U^\alpha(g^2)]\notag\\
&&=\frac{1}{|\mathcal{G}|}\sum_{g\in\mathcal{G}}\sum_\sigma\sum_{h\in\mathcal{G}_{\vec{k}_\sigma}}\delta_{g^2,h}z_{g,g}\chi^\alpha_{\vec{k}_\sigma}(h)\notag\\
&&=\frac{1}{|\mathcal{G}_{\vec{k}}|}\sum_{g\in\mathcal{G}}z_{g,g}\chi^\alpha_{\vec{k}}(g^2)\notag\\
&&=\frac{1}{|\mathcal{G}_{\vec{k}}/T|}\sum_{g\in\mathcal{G}/T}\delta_{p_g\vec{k},-\vec{k}}''\,z_{g,g}\chi^\alpha_{\vec{k}}(g^2)
\end{eqnarray}
To go to the third line, we used the fact that contributions from each star of $\vec{k}$ are identical and replaced $\sum_\sigma$ by the factor $|\mathcal{G}|/|\mathcal{G}_{\vec{k}}|$. To go to the last line, we performed the sum over the translation subgroup $T$, which results in the constraints $\sum_{\vec{t}_{\vec{R}}\in T}e^{-i\vec{k}\cdot(p_g\vec{t}_{\vec{R}}+\vec{t}_{\vec{R}})}=|T|\delta_{p_g\vec{k},-\vec{k}}''$. Therefore, the TR index can be computed given by~\cite{Bradley}
\begin{eqnarray}
&&\eta_{\mathcal{T}}\frac{1}{|\mathcal{G}_{\vec{k}}/T|}\sum_{g\in\mathcal{G}/T}\delta_{p_g\vec{k},-\vec{k}}''z_{g,g}\chi^\alpha_{\vec{k}}(g^2)\notag\\
&&=\begin{cases}
+1& \text{: Degeneracy is unchanged.}\\
-1& \text{: Two $u^\alpha_{\vec{k}}$'s are paired under TR.}\\
0 & \text{: $u^\alpha_{\vec{k}}$ is paired with another irrep $u^\beta_{\vec{k}}$}.
\end{cases}
\end{eqnarray}
As explained in the main text, when an irrep $u^{\alpha}_{\vec k}$ is paired with a different irrep $u^{\beta}_{\vec k}$ under TR, we simply add to $\mathcal C$ an additional compatibility relation $n_{\vec k}^{\alpha} = n_{\vec k}^{\beta}$; when $u^{\alpha}_{\vec k}$ is paired with itself, we demand $n_{\vec k}^{\alpha}$ to be an even integer, which can be achieved by redefining ${\tilde n}_{\vec k}^{\alpha} \equiv  n_{\vec k}^{\alpha}/2$ and a corresponding rewriting of $\mathcal C$ in terms of $\tilde {\vec n}$.

\section{Atomic Insulators as Band Structures
\label{app:AI}}

In \ref{app:SymBS}, we characterized each BS by $\vec{n}$, the set of integers $n^{\alpha}_{\vec k}$ that specifies the representation contents of the BS.  In this section, we derive a general formula that gives $\vec{n}$ for each AI.  


\subsection{Wyckoff position and site symmetry representations}
An AI is specified by the location of the sites on which atomic orbitals sit, and the type of orbitals on each site.  Mathematically, these two inputs correspond to the choice of a Wyckoff position and the representation of the site symmetry group of a site in the Wyckoff position.

Just as we defined the little group of $\vec{k}$, let us define the little group of $\vec{x}$ as the subgroup of $\mathcal G$ that leaves $\vec{x}$ invariant.  We call it the site symmetry group $\mathcal G_{\vec x}$ of $\vec{x}$.
As a set, we have
\begin{equation}
\mathcal{G}_{\vec x}=\{h\in \mathcal{G}~|~h(\vec{x})\equiv p_h\vec{x}+\vec{t}_h=\vec{x}\}.
\end{equation}
Points in real space are classified based on their little group. Namely, two points $\vec{x}_1$ and $\vec{x}_2$ belong to the same Wyckoff position iff there exists $g\in\mathcal{G}$ such that $\mathcal{G}_{\vec{x}_2}=g\mathcal{G}_{\vec{x}_1}g^{-1}$. The full list of different Wyckoff positions (in the real space) is available in Ref.~\cite{ITC}.

Let us pick a site $\vec{x}$ in the unit cell $\text{UC}$.  By definition, elements of $\mathcal{G}$ not belonging to $\mathcal{G}_{\vec x}$ will move $\vec{x}$.  The crystallographic orbit $\{g(\vec{x})|g\in\mathcal{G}\}$ defines a $\mathcal{G}$-symmetric lattice $L_{\vec{x}}$. Let $\{\vec{x}_\sigma\}_{\sigma=1,2,\ldots}$ ($\vec{x}_1\equiv\vec{x}$) be the lattice points in $\text{UC}$.   We choose $\{g_\sigma\}_{\sigma=1,2,\ldots}$ from $\mathcal{G}$ in such a way that $g_{\sigma=1}=e$ ($e\in\mathcal{G}$ is the identity) and $g_\sigma(\vec{x})=\vec{x}_\sigma$ for $\sigma=2,3,\ldots$. Namely, $\{g_\sigma\}_{\sigma=1,2,\ldots}$ is a complete set of representatives of $\mathcal{W}_{\vec{x}}\equiv(\mathcal{G}/\mathcal{G}_{\vec{x}})/T$.

We want to introduce an orbital on every site of $L_{\vec{x}}$ in a symmetric manner.  To that end, 
let us first put states $\{|\phi_{\vec{x},i,\vec{k}}^r\rangle\}_{i=1}^{\text{dim}[u_{\vec{x}}^r]}$ on $\vec{x}$ that obey an irrep $u_{\vec{x}}^r$ of $\mathcal{G}_{\vec{x}}$:
\begin{eqnarray}
\hat{h}|\phi_{\vec{x},i,\vec{k}}^r\rangle&=&\sum_j|\phi_{\vec{x},j,p_h\vec{k}}^r\rangle [u_{\vec{x}}^r(h)]_{ji},\\
u_{\vec{x}}^r(h)u_{\vec{x}}^r(h')&=&z_{h,h'}u_{\vec{x}}^r(hh'),\\
\hat{h}(\hat{h}'|\phi_{\vec{x},i,\vec{k}}^r\rangle)&=&z_{h,h'}\hat{(hh')}|\phi_{\vec{x},i,\vec{k}}^r\rangle
\end{eqnarray}
for $h, h'\in \mathcal{G}_{\vec{x}}$ and
\begin{eqnarray}
\hat{t}_{\vec{R}}|\phi_{\vec{x},i,\vec{k}}^r\rangle=|\phi_{\vec{x},i,\vec{k}}^r\rangle e^{-i\vec{k}\cdot\vec{R}}
\end{eqnarray}
for $t_{\vec{R}}\in T$.  The orbitals at other sites of $L_{\vec{x}}$ are then defined by $|\phi_{\vec{x}_\sigma,i,\vec{k}}^r\rangle\equiv\hat{g}_\sigma |\phi_{\vec{x},i,p_\sigma^{-1}\vec{k}}^r\rangle$.  As before, $z_{g,g'}=\pm1$ is a factor system of the projective representation originating from the spin degrees of freedom. 

The choice of the position $\vec{x}$ to start with and the choice of an irrep $u_{\vec{x}}^r$ of the site symmetry group $\mathcal{G}_{\vec{x}}$ will specify an AI and its representation contents as we will see now.

\subsection{Representation contents of an AI}
To determine the transformation of $\{|\phi_{\vec{x}_\sigma,i,\vec{k}}^r\rangle\}_{\sigma,i,\vec{k}}$ under $\mathcal{G}$, note that any element $g\in\mathcal{G}$ can be uniquely decomposed as $g=t_{\vec{R}} g_\sigma h$ where $t_{\vec{R}}\in T$ and $h\in \mathcal{G}_{\vec{x}}$. In particular, we can decompose $gg_\sigma$ as $t_{\vec{R}}g_{\sigma'}h$ with $\vec{R}=g(\vec{x}_{\sigma})-\vec{x}_{\sigma'}$. Therefore,
\begin{eqnarray}
&&\hat{g}|\phi_{\vec{x}_\sigma,i,\vec{k}}^r\rangle\notag\\
&&=z_{g,g_{\sigma}}\hat{(gg_\sigma)}|\phi_{\vec{x},i,p_\sigma^{-1}\vec{k}}^r\rangle\notag\\
&&=z_{g,g_{\sigma}}\hat{(t_{\vec{R}}g_{\sigma'}h)}|\phi_{\vec{x},i,p_\sigma^{-1}\vec{k}}^r\rangle\notag\\
&&=\frac{z_{g,g_{\sigma}}}{z_{g_{\sigma'},h}}\hat{t}_{\vec{R}}\hat{g}_{\sigma'}(\hat{h}|\phi_{\vec{x},i,p_\sigma^{-1}\vec{k}}^r\rangle)\notag\\
&&=\frac{z_{g,g_{\sigma}}}{z_{g_{\sigma'},h}}\sum_{i'}(\hat{t}_{\vec{R}}|\phi_{\vec{x}_{\sigma'},i',p_g\vec{k}}^r\rangle) [u_{\vec{x}}^r(h)]_{i'i}\notag\\
&&=\sum_{\sigma',i',\vec{k}'}|\phi_{\vec{x}_{\sigma'},i',\vec{k}'}^r\rangle[U_{\vec{x}}^r(g)]_{\sigma'i'\vec{k}',\sigma i\vec{k}},
\end{eqnarray}
where
\begin{eqnarray}
\label{eq:induced2}
&&[U_{\vec{x}}^r(g)]_{\sigma'i'\vec{k}',\sigma i\vec{k}}\\
&&=\delta'_{\vec{x}_{\sigma'}, g(\vec{x}_{\sigma})}\delta''_{\vec{k}',p_g\vec{k}}e^{-i\vec{k}'\cdot(g(\vec{x}_{\sigma})-\vec{x}_{\sigma'})}
\frac{z_{g,g_{\sigma}}}{z_{g_{\sigma'},h_{\sigma',\sigma}^g}}[u_{\vec{x}}^r(
h_{\sigma',\sigma}^g
)]_{i'i},\notag
\end{eqnarray}
$\delta'_{\vec{x}_1,\vec{x}_2}=1$ only when $\vec{x}_1=\vec{x}_2$ modulo a lattice vector, and
\begin{eqnarray}
h_{\sigma',\sigma}^g\equiv g_{\sigma'}^{-1}t_{\vec{x}_{\sigma'}-g(\vec{x}_{\sigma})}gg_\sigma. 
\end{eqnarray}
Note that $h_{\sigma',\sigma}^g\in\mathcal{G}_{\vec{x}}$ when $\delta'_{\vec{x}_{\sigma'}, g(\vec{x}_{\sigma})}=1$.  An AI constructed from an irrep $u_{\vec{x}}^r$ of $\mathcal{G}_{\vec{x}}$ has a representation $U_{\vec{x}}^r$ of $\mathcal{G}$.   Although the discussion here is similar to the derivation of the induced representation $U^\alpha$ of $\mathcal{G}$ starting from an irrep $u_{\vec{k}}^\alpha$ of $\mathcal{G}_{\vec{k}}$ in \ref{app:SymBS}, there is an important difference. That is, the representation $U_{\vec{x}}^r$ is in general reducible, unlike $U^\alpha$ which is always irreducible whenever $u_{\vec{k}}^\alpha$ is irreducible.  

Let us focus on a particular $\vec{k}$ in BZ.  The AI's representation of $\mathcal{G}_{\vec{k}}$ is immediately given by $U_{\vec{x}}^r$ by restricting $\mathcal{G}$ to $\mathcal{G}_{\vec{k}}$. In particular, its character is 
\begin{eqnarray}
&&\chi_{\vec{x},\vec{k}}^r(g)\equiv \text{tr}[U_{\vec{x}}^r(g)]\notag\\
&&=\sum_{\sigma=1}^{|\mathcal{W}_{\vec{x}}|}\delta'_{\vec{x}_{\sigma}, g(\vec{x}_{\sigma})}\,e^{-i\vec{k}\cdot(g(\vec{x}_{\sigma})-\vec{x}_{\sigma})}\frac{z_{g,g_{\sigma}}}{z_{g_{\sigma},h_{\sigma,\sigma}^g}}\chi_{\vec{x}}^r(h_{\sigma,\sigma}^g),
\label{appAIchi}
\end{eqnarray}
where $\chi_{\vec{x}}^r(h)\equiv\text{tr}[u_{\vec{x}}^r(h)]$.  Note that this result does not depend on the choice of $\vec{x}_\sigma$; even if one chose $\vec{x}_{\sigma}'=\vec{x}_{\sigma}+\vec{R}_\sigma$ instead, the character is unchanged.  

Let $\chi_{\vec{k}}^\alpha(g)=\text{tr}[u_{\vec{k}}^\alpha(g)]$ be the character of an irrep $u_{\vec{k}}^\alpha$ of $\mathcal{G}_{\vec{k}}$. Then, the irrep $u_{\vec{k}}^\alpha$ appears in $U_{\vec{x}}^r(g)$
\begin{eqnarray}
n_{\vec{k}}^{\alpha}=\sum_{g\in \mathcal{G}_{\vec{k}}/T}\frac{1}{|\mathcal{G}_{\vec{k}}/T|}[\chi_{\vec{k}}^{\alpha}(g)]^*\chi_{\vec{x},\vec{k}}^r(g)\in\mathbb{Z}_{\geq0}
\end{eqnarray}
times. This is the formula that gives $\vec{n}$ for a AI in general.

\subsection{The general position}
As a special case of the above general discussion, let us assume that the site symmetry group $\mathcal{G}_{\vec{x}}$ is trivial, i.e., $\mathcal{G}_{\vec{x}}=\{e\}$. In other words, let us assume that $\vec{x}$ belongs to the general Wyckoff position.  There is only one trivial representation $u_{\vec{x}}^{r=1}(e)=1$ for such a generic position.  
In this case, Eq.~\eqref{appAIchi} reduces to
\begin{eqnarray}
&&\chi_{\vec{x},\vec{k}}^{r=1}(g)=|\mathcal{G}/T|\delta_{g,e}.
\end{eqnarray}
Therefore,  using $\chi_{\vec{k}}^\alpha(e)=\text{tr}[u_{\vec{k}}^\alpha(e)]=\text{dim}[u_{\vec{k}}^\alpha]$, we get 
\begin{equation}
n_{\vec{k}}^{\alpha}|_{\text{generic position}}=\frac{|\mathcal{G}/T|}{|\mathcal{G}_{\vec{k}}/T|}\text{dim}[u_{\vec{k}}^\alpha]\geq1.
\label{app:genericAI}
\end{equation}
Namely, the AI constructed from the trivial orbital on a generic position $\vec{x}$ contains every irrep $u_{\vec{k}}^\alpha$ at least once at each $\vec{k}$.

\subsection{The special position}
Next, consider a position $\vec{x}$ with a nontrivial $\mathcal{G}_{\vec{x}}>{e}$.  In other words, $\vec{x}$ belongs to a special Wyckoff position.  In this case, there are several irreps $u_{\vec{x}}^r$. We have a sum-rule separately for each $\vec{x}$:
\begin{equation}
n_{\vec{k}}^{\alpha}|_{\text{generic position}}=\sum_{r:\text{all irreps on $\vec{x}$}} \text{dim}[u_{\vec{x}}^r]\,\,n_{\vec{k}}^{\alpha}|_{\text{irrep $u_{\vec{x}}^r$ on $\vec{x}$}}.
\end{equation}

\subsection{TR invariant AIs}
To construct a TR invariant AI, we have to determine if the time-reversal (TR) symmetry $\mathcal{T}$ enhances the degeneracy of an irrep $u_{\vec{x}}^r$.  This can be easily done by the method reviewed in \ref{app:SymBS}.
Namely, one should compute the following sum, which can be either $+1$, $0$, or $-1$. 
\begin{eqnarray}
&&\eta_{\mathcal{T}}\frac{1}{|\mathcal{G}_{\vec{x}}|}\sum_{h\in \mathcal{G}_{\vec{x}}}z_{h,h}\,\chi_{\vec{x}}^r(h^2)\notag\\
&&=
\begin{cases}
+1& \text{: $u_{\vec{x}}^r$ by itself is TR invariant.}\\
-1& \text{: Two $u_{\vec{x}}^r$'s are paired under TR.}\\
0 & \text{: $u_{\vec{x}}^r$ and $(u_{\vec{x}}^r)^*$ are different and are paired}.
\end{cases}
\end{eqnarray}
Here, $\chi_{\vec{x}}^r(h)\equiv\text{tr}[u_{\vec{x}}^r(h)]$ and $\eta_{\mathcal{T}}=-1$ ($\eta_{\mathcal{T}}=+1$) for the spinful (spinless) fermions.   If the sum is either $-1$ or $0$, the AI constructed from the irrep $u_{\vec{x}}^r$ on the site $\vec{x}$ alone is not TR symmetric, and one has to make a proper stacking with its TR pair.

\subsection{Uniform basis}
So far we have presented two constructions of a representation of $\mathcal{G}$: In \ref{app:SymBS} we constructed one from an irrep $u_{\vec{k}}^\alpha$ of $\mathcal{G}_{\vec{k}}$, and 
the present note gives another one from an irrep $u_{\vec{x}}^r(h)$ of $\mathcal{G}_{\vec{x}}$.   There is yet another construction of a representation of $\mathcal{G}$. Suppose that we know a representation $v(p_g)$ of the point group $\mathcal{G}/T$.  Then a representation of $\mathcal{G}$ is given by
\begin{eqnarray}
&&[U(g)]_{\sigma'i'\vec{k}',\sigma i\vec{k}}\notag\\
&&=\delta'_{\vec{x}_{\sigma'}, g(\vec{x}_{\sigma})}\,\delta''_{\vec{k}',p_g\vec{k}}e^{-i\vec{k}'\cdot(g(\vec{x}_{\sigma})-\vec{x}_{\sigma'})}[v(p_g)]_{i'i}.
\end{eqnarray}
The advantage of this representation is that we have the same representation $v(p_g)$ on every site. As a result, the effect of spin-orbit coupling, for example, can be interpreted much easily than $U_{\vec{x}}^r$.  Our leSM example discussed in the main text is formulated using this representation. 
The drawback is that this construction generally requires as input particular representations of $\mathcal G_{\vec x}$, in contrast to the earlier construction which applies to any representation, in particular to the irreducible ones.

\section{Structure of $\{ {\rm BS} \}$ and Computation of $X_{\rm BS}$ 
\label{app:X_BS}}
\subsection{Mathematical details}
Here, we discuss some relatively formal aspects of our mathematical framework. We will start with the claim
\begin{equation}\begin{split}\label{eq:}
\{ {\rm BS} \} \equiv \ker \mathcal C \cap \mathbb Z^D ~\Rightarrow ~ \{ {\rm BS} \} \simeq \mathbb Z^{d_{\rm BS}},
\end{split}\end{equation}
where $d_{\rm BS} \equiv \dim \ker \mathcal C$.
This claim can be interpreted geometrically: Embedded in $\mathbb R^D$, $\mathbb Z^{D}$ is a hypercubic lattice and $\ker \mathcal C$ defines a $d_{\rm BS}$-dimensional hyperplane. $\{ {\rm BS}\}$ is then the collection of lattice sites sliced by $\ker \mathcal C$, which defines a Bravais lattice in $d_{\rm BS}$ dimensions.

Alternatively, this can also be understood algebraically, as we now discuss in details. First observe $\{ {\rm BS} \} \leq \mathbb Z^D$. Since any subgroup of a finitely generated abelian group is again finitely generated, and that no element in $\mathbb Z^D$ has finite order, we see that $\{ {\rm BS} \} \simeq \mathbb Z^{d}$ for some $d\leq D$. 
Next, note that as $\mathcal C$ is a matrix of integer coefficients, its solution space can be identified as a vector subspace of $\mathbb Q^{D}$ (instead of $\mathbb R^D$). Let $\{ \vec q_i ~:~ i = 1,\dots, d_{\rm BS} \}$  be any complete basis for $\ker \mathcal C \simeq \mathbb Q^{d_{\rm BS}}$. Since only a finite number of rational numbers are involved, we can always multiply the basis by the least-common multiple of all the denominators to arrive at an integer-valued basis. This implies $\{ {\rm BS} \}$ has at least $d_{\rm BS}$ linearly-independent elements, i.e.~$d\geq d_{\rm BS}$. 
Now also observe that as $\{ {\rm BS} \} \leq \ker \mathcal C$, $\ker \mathcal C$ has at least $d$ linearly-independent vectors, which implies $ d_{\rm BS} \leq d \leq d_{\rm BS}$.

Next we discuss the computation of $X_{\rm BS} \equiv \{ {\rm BS} \} / \{ {\rm AI} \}$, where $\{ {\rm AI} \} \simeq \mathbb Z^{d_{\rm AI}} \leq \{ {\rm BS} \}$ denotes the subgroup of BS arising from AIs. The first step in the analysis is to compare their ranks. $d_{\rm BS}$, a property of $\mathcal C$, is determined once all compatibility relations are found. $d_{\rm AI}$ can be computed as follows: Any AI can be understood as a stack of those arising from fully occupying an irrep of the site-symmetry group of a Wyckoff position. Hence, by focusing on the finite number of AIs arising from these irreps, we can find a (generally over-complete) basis for $\{{\rm AI} \}$ using the formalism developed in \ref{app:SymBS} and \ref{app:AI}. Finally we simply extract the number of linearly independent combinations among them, which is by definition $d_{\rm AI}$. 

Generally, we have $d_{\rm AI} \leq d_{\rm BS}$, and the general structure of the quotient group is given by 
\begin{equation}\begin{split}\label{eq:X_gen}
X_{\rm BS} = \mathbb Z^{d_{\rm BS}- d_{\rm AI}} \times \mathbb Z_{s_1} \times \mathbb Z_{s_2} \times \dots \times \mathbb Z_{s_{d_{\rm AI}}}.
\end{split}\end{equation}
It remains to compute the integers $s_i\geq 1$.
In more physical terms, any element of $X_{\rm BS}$ corresponds to a class of BSs that cannot be obtained from integer combinations (i.e.~stacking) of entries in $\{ {\rm AI}\}$. Now consider a $\vec b \in \{ {\rm BS} \}$ with its equivalence class $[\vec b]$ being the generator of a factor $Z_{s_i}$ in $X_{\rm BS}$. From definition, $s_i [\vec b] = [ s_i \vec b] $ is the trivial element of $X_{\rm BS}$, i.e.~$s_i \vec b$ is an AI. Running the argument in reverse, the torsion (i.e.~finite-order) elements of $X_{\rm BS}$ correspond to (the classes of) fractions of AIs that are nonetheless in $\{ {\rm BS}\}$, and hence to compute the $s_i$'s in Eq.~\eqref{eq:X_gen} one simply studies the possible set of coefficients for which $\sum_{i=1}^{d_{\rm AI}} q_i \vec a_i$, $q_i \in \mathbb Q$, is integer-valued. This can be readily computed using the Smith normal form, which is, loosely speaking, an integer-valued version of the singular-value decomposition (in our context).

An interesting observation is that, for all the $230 \times 4$ cases we studied, corresponding to all the SGs assuming spinless or spinful fermions with or without TR symmetry, we found $d_{\rm BS} = d_{\rm AI}$. This implies a basis for $\{{\rm BS }\}$ can be obtained by a suitable combination of the basis of $\{ {\rm AI} \}$ using rational coefficients, which are found in the computation of $s_i$ described above. Equivalently, we found that a BS can always be expanded as 
\begin{equation}\begin{split}\label{eq:BS_AI_exp}
{\rm BS} = \sum_{\vec{x}} \sum_{r} q_{\vec{x},r} \, \vec{n}_{\vec{x}, r},
\end{split}\end{equation}
where  $\vec{n}_{\vec{x}, r}$ denotes the $\vec{n}$ of the AI arising from fully occupying a site-symmetry group irrep $u_{\vec{x}}^r$ of the position $\vec{x}$ (it is sufficient to choose one representative of $\vec{x}$ for each Wyckoff position), and $q_{\vec{x},r}\in\mathbb{Q}$ are rational numbers. This is very similar to the decomposition of energy bands at high-symmetry momenta into irreps of the little group, except that this is now performed in a global manner over the BZ. An interesting open question is whether there is any symmetry setting, say when one studies the remaining magnetic SGs, for which $d_{\rm BS} > d_{\rm AI}$ and therefore leads to an infinite $X_{\rm BS}$ (i.e.~some BSs remain non-atomic no matter how many copies we take). Alternatively, if such equality always holds for any symmetry settings, there should be a more elegant method to prove that $X_{\rm BS}$ is always finite.

In closing, we comment that the expansion in Eq.~\eqref{eq:BS_AI_exp} is similar in spirit to the notion of elementary energy bands developed in a series of work by Refs.~\cite{Zak1999,Zak2000,Zak2001}. In our language, these earlier results focus on such a decomposition between the AIs, and whether or not the building blocks of such decomposition, dubbed elementary energy bands, can be split into energy bands of lower fillings.
These earlier works approached the problem from a real-space perspective, which in our language is about the structure of $\{ {\rm AI}\}$. In contrast, our formalism, centered on the structure of $\{ {\rm BS} \}$, automatically captures all the momentum-space constraints and is more suited for studying topological band structures: our notion of a global decomposition allows for quantum interference in momentum space, which is more general than that in Refs.~\cite{Zak1999,Zak2000,Zak2001}. Finally, we also remark that the results concerning energy band connectivity in Refs.~\cite{Zak1999,Zak2000,Zak2001}. has interesting implications on finding leSMs (specifically, for spinless fermions with TR symmetry).

(Note: From a more mathematical perspective, the earlier results in Refs.~\cite{Zak1999,Zak2000,Zak2001} are concerned with a decomposition in terms of the direct sum $\oplus$, whereas we have first made a generalization $\oplus \rightarrow +$, akin to how representation rings are constructed, and then further assert that the decomposition is physically meaningful with rational coefficients, as long as the resulting (formal) sum lies in $\{ {\rm BS} \}$.)

\subsection{Physical relevance of $X_{\rm BS}$}
Having discussed the mathematical aspect of the formalism, we now comment on why such notions are relevant to physical band structures. 
Recall we have motivated the definition of $\{ {\rm BS} \}$ by asserting that, as long as only symmetry properties are concerned, a set of energy bands can be labelled simply by a count of the multiplicities of each irrep at the high-symmetry momenta. A priori, it is insensible to say that an irrep appears a negative number of times. As mentioned in the main text, this leads to an additional condition in connecting the entries of $\{ {\rm BS} \}$ to physical band structures--for $\vec n \in \{ {\rm BS} \}$ to be physical, all components of $\vec n$ must be non-negative. Equivalently, one can define a physical subset
\begin{equation}\begin{split}\label{eq:}
\{ {\rm BS} \}_{\rm P} \equiv \{ {\rm BS} \} \cap \mathbb Z^{D}_{\geq 0} \subset \{ {\rm BS} \}.
\end{split}\end{equation}
As we have explained, all elements in $\{ {\rm BS} \}_{\rm P}$ enjoy the properties of a general element in $\{ {\rm BS} \}$, say the decomposition in Eq.~\eqref{eq:BS_AI_exp}. Therefore, the inclusion of unphysical entires, crucial for the group properties of $\{ {\rm BS} \}$, should be viewed merely as a mathematical way to simplify the analysis of the physical problem.
(If one insists, one can identify $\{ {\rm BS} \}_{\rm P}$ as a commutative monoid enjoying similar properties as the group $\{ {\rm BS} \}$. However, we do not find this perspective particularly useful in our discussion, and would rather stick with $\{ {\rm BS} \}$, a simpler mathematical gadget.)

However, there are still two points we have to establish in order in order to quantify the relevance of our mathematical framework to the study of real, physical band structures. First, we argue that all entries of $\{ {\rm BS} \}_{\rm P}$ correspond to physically realizable band structures.
To see this, consider an arbitrary element of $\{ {\rm BS} \}_{\rm P}$, and let a set of physical bands possess the desired irrep content at all high-symmetry momentum points. (If only high-symmetry lines or planes are present, we choose arbitrary, isolated representatives.)
This is always possible by a suitable arrangement of the energies of the irreps at these isolated momentum points. By symmetries and continuity, all compatibility relations are locally satisfied near these points. Suppose there are still certain band crossings obstructing us from identifying this set of bands as a BS. Since any pair of bands carrying the same symmetry representation will generically anti-cross, these band crossings must correspond to an exchange of irreps between our target set of bands and the others. As all compatibility relations are globally satisfied by assumption, such exchange must be accidental in nature, i.e.~it is possible to perturb the Hamiltonian in a symmetric fashion to get rid of all the band crossings. This then leads to a BS corresponding to the specified element in  $\{ {\rm BS} \}_{\rm P}$.

Note that, however, there is a subtlety in the statement on generic anti-crossing: Topological band degeneracies, like Weyl points, can only be pushed away but not lifted, unless their topological charges are neutralized by their partners. 
This does not concern us, since by our definition of BS we will only be interested in band gaps at high-symmetry momenta, and therefore as long as these degeneracies can be moved away they do not affect our discussion. This also explains why the notion of reSM, which cannot be insulating due to the specified representation content, is still consistent with our notion of BS.

Second, we show that all nontrivial classes in $X_{\rm BS}$ have physical representatives.
Let $\vec b$ be a representative of a nontrivial class in $X_{\rm BS}$ which is not physical, i.e.~certain entries in $\vec b $ are negative. Now, we consider a small corollary from \ref{app:AI}: all irreps appear at least once in the AI corresponding to the generic Wyckoff position [see Eq.~\eqref{app:genericAI}]. Therefore, we can always stack $\vec b$ with a sufficiently large number of copies of the generic AI and rectify the representation content. By definition, such stacking leads to a physical BS belonging to the same nontrivial class as $\vec b$, and hence all classes of $X_{\rm BS}$ have physical representatives.

\subsection{Lower dimensional systems}
Here we discuss  $X_{\rm BS}$ for (quasi-)1D and 2D systems.  As far as spinful electrons are concerned, the fact that lower dimensional systems in reality are embedded in the 3D space cannot be neglected, since the electronic spin degree of freedom is a projective representation of $O(3)$, the rotation of the 3D space.  For brevity we will focus on 2D systems but the 1D case can be discussed in the same way.

The symmetry groups for 2D lattices lying in the 3D space are called layer groups.  An element $h$ of a layer group $\mathcal{L}$ maps $(x,y,z)$ to $(x',y',z')$, where $(x',y')=q_h(x,y)+\vec{s}_h$ ($q_h$ is a $O(2)$ matrix and $\vec{s}_h$ is a two component vector) and $z'=\xi_hz$ ($\xi_h=\pm 1$).  The translation subgroup $T_{\text{2D}}$ of $\mathcal{L}$ is a group of lattice translations in the 2D plane, giving rise to the 2D crystal momentum $(k_x,k_y)$.  (At this moment $k_z$ is not defined.)  One can follow the same steps as in 3D to define $\{\rm BS^{\mathcal{L}}\}$ and $\{\rm AI^{\mathcal{L}}\}$, and $X_{\rm BS}^{\mathcal{L}}=\{\rm BS^{\mathcal{L}}\}/\{\rm AI^{\mathcal{L}}\}$ for a layer group $\mathcal{L}$.

To make use of our results established for space groups, let us consider the space group $\mathcal{G}$ corresponding to a layer group $\mathcal{L}$, which is simply the layer group $\mathcal{L}$ endowed with a lattice translation $T_{z}$ along $z$.  More precisely, $\mathcal{G}$ is given by the semi-direct product of $\mathcal{L}$ and $T_{z}$, whose element $g=(h,t)\in\mathcal{G}$ ($h\in\mathcal{L}$ and $t\in T_{z}$) maps $(x,y,z)$ to $(x',y',z')$, where $(x',y')=q_h(x,y)+\vec{s}_h$ and $z'=\xi_h z+t$.  The product of $g=(h,t)\in\mathcal{G}$ and $g'=(h',t')\in\mathcal{G}$ is defined as $gg'=(hh',t+\xi_h t')$.  We list the corresponding space group $\mathcal{G}$ for each layer group $\mathcal{L}$ in Table~\ref{tab:LGSG}. 
Note that there is one different entry when our table is compared to the one provided in Ref.~\cite{LGTab}: we found that the correct correspondence for layer group 35 should be space group 38.
To avoid confusion, we use the notations $\{\rm BS^{\mathcal{G}}\}$, $\{\rm AI^{\mathcal{G}}\}$, and $X_{\rm BS}^{\mathcal{G}}=\{\rm BS^{\mathcal{G}}\}/\{\rm AI^{\mathcal{G}}\}$ for a space group $\mathcal{G}$ in this section.

Given a layer group $\mathcal L$ and the corresponding space group $\mathcal G$, we expect $X_{\rm BS}^{\mathcal G}$ to encapsulate that of $X_{\rm BS}^{\mathcal L}$: stacking lower-dimensional nontrivial phases by translation symmetries will naturally give rise to weak topological phases, which are nontrivial as long as the translation symmetries remain intact, i.e., a nontrivial BS of $\mathcal L$ will never become trivial when lifted to $\mathcal G$ upon stacking. Generally, we also expect  $X_{\rm BS}^{\mathcal G}$  to be richer than  $X_{\rm BS}^{\mathcal L}$, since certain strong  phases should become possible. These observations can be summarized by asserting the subgroup relation $X_{\rm BS}^{\mathcal L} \leq X_{\rm BS}^{\mathcal G}$(we will explain shortly the precise meaning of this symbolic relation). In the following, we formalize these observations and provide a (technical) proof for this relation. Before we dwell into the technical details, we remark that, given this natural subgroup relation, the finiteness of $X_{\rm BS}^{\mathcal G}$ implies that of $X_{\rm BS}^{\mathcal L}$, and therefore $X_{\rm BS}^{\mathcal L}$ can be readily computed using only the data on AI without finding and solving the compatibility relations.

Let us introduce a group homomorphism $f: \{\rm BS^{\mathcal{L}}\}\rightarrow\{\rm BS^{\mathcal{G}}\}$ through stacking of layers. Namely, starting from a given $\vec{b}\in \{\rm BS^{\mathcal{L}}\}$ of a layer group $\mathcal{L}$, we can get a band structure of $\mathcal{G}$ by stacking infinite copies of $\vec{b}$ in the $z$ direction.  To make this idea more concrete, let us take a tight-binding model $\hat{H}_{\mathcal{L}}$ symmetric under $\mathcal{L}$ such that the lowest $\nu$ bands of $\hat{H}_{\mathcal{L}}$ are isolated from other bands by a band gap at all high-symmetry points of the 2D BZ, and  the combination of irreps of the lowest $\nu$ bands precisely agrees with $\vec{b}$.  Given $\hat{H}_{\mathcal{L}}$, one can generate a $\mathcal{G}$ symmetric tight-binding model by
\begin{equation}
\hat{H}_{\mathcal{G}}=\sum_{t \in T_{z}}\hat{t}\hat{H}_{\mathcal{L}}\hat{t}^{-1}.
\end{equation}
Since there is no inter-layer hopping, the band structure of $\hat{H}_{\mathcal{G}}$ is completely flat as a function of $k_z$, and as a result, the lowest $\nu$ bands remain isolated from other bands by a band gap at all high-symmetry points of the 3D BZ.  This band structure of $\hat{H}_{\mathcal{G}}$ defines $f(\vec{b})\in\{\rm BS^{\mathcal{G}}\}$.  By construction, the homeomorphism $f(\vec{b}_1+\vec{b}_2)=f(\vec{b}_1)+f(\vec{b}_2)$ is obvious. Furthermore, $f$ maps $\vec{a}\in\{\rm AI^{\mathcal{L}}\}$ to $f(\vec{a})\in\{\rm AI^{\mathcal{G}}\}$. Therefore, $f$ defines a group homomorphism
\begin{equation}
f: X_{\rm BS}^{\mathcal{L}}\rightarrow X_{\rm BS}^{\mathcal{G}}
\label{app:fX}
\end{equation}
Below we show that this $f$ is injective.  If this is the case, $\tilde{X}_{\rm BS}^{\mathcal{L}}\equiv\text{Im}f$ is isomorphic to $X_{\rm BS}^{\mathcal{L}}$ and is a subgroup of $X_{\rm BS}^{\mathcal{G}}$.

To this end, let us introduce a projection $p$ that defines a homomorphism from $\{\rm BS^{\mathcal{G}}\}$ to $\{\rm BS^{\mathcal{L}}\}$. Given $\vec{B}\in\{\rm BS^{\mathcal{G}}\}$, one can project out all entries of $\vec{B}$ associated with high-symmetry points with $k_z\neq0$, keeping only entries associated with high-symmetry points with $k_z=0$. By definition $p(\vec{B})$ satisfies all compatibility conditions imposed on the irreps of $\mathcal{L}$ and hence is an element of $\{\rm BS^{\mathcal{L}}\}$.  The projection $p$ in fact acts as an inverse of $f$: $p(f(\vec{b}))=\vec{b}$.  Furthermore, $p$ maps $\vec{A}\in\{\rm AI^{\mathcal{G}}\}$ to $p(\vec{A})\in\{\rm AI^{\mathcal{L}}\}$.  This can be seen by the fact that, for every Wyckoff position of $\mathcal{G}$, there exists a Wyckoff position of $\mathcal{L}$ that is either identical, or differs only by the value of $z$.  The difference of $z$ does not affect $p(\vec{A})$ since the projection $p$ sets $k_z=0$.  Given these properties of $p$, it is easy to prove that $\text{ker}f=\{e\}$.  If not, there must be a $\vec{b}\in\{\rm BS^{\mathcal{L}}\}$ belonging to a nontrivial class of $X_{\rm BS}^{\mathcal{L}}$ which is mapped to $\vec{A}\in\{\rm AI^{\mathcal{G}}\}$ by $f$. Then $\vec{b}=p(\vec{A})\in\{\rm AI^{\mathcal{L}}\}$ is an AI, contradicting with the assumption that $\vec{b}$ is nontrivial. Hence the proof.

\begin{table}[h]
\begin{center}
\caption{\bf Correspondence between space groups and layer groups.
\label{tab:LGSG}}
\begin{tabular}{cc|cc|cc|cc} \hline \hline
~~LG~~& ~~SG~~& ~~LG~~& ~~SG~~& ~~LG~~& ~~SG~~& ~~LG~~& ~~SG~~\\
\hline
{1} & {1} & {21} & {18} & {41} & {51} & {61} & {123}\\
{2} & {2} & {22} & {21} & {42} & {53} & {62} & {125}\\
{3} & {3} & {23} & {25} & {43} & {54} & {63} & {127}\\
{4} & {6} & {24} & {28} & {44} & {55} & {64} & {129}\\
{5} & {7} & {25} & {32} & {45} & {57} & {65} & {143}\\
{6} & {10} & {26} & {35} & {46} & {59} & {66} & {147}\\
{7} & {13} & {27} & {25} & {47} & {65} & {67} & {149}\\
{8} & {3} & {28} & {26} & {48} & {67} & {68} & {150}\\
{9} & {4} & {29} & {26} & {49} & {75} & {69} & {156}\\
{10} & {5} & {30} & {27} & {50} & {81} & {70} & {157}\\
{11} & {6} & {31} & {28} & {51} & {83} & {71} & {162}\\
{12} & {7} & {32} & {31} & {52} & {85} & {72} & {164}\\
{13} & {8} & {33} & {29} & {53} & {89} & {73} & {168}\\
{14} & {10} & {34} & {30} & {54} & {90} & {74} & {174}\\
{15} & {11} & {35} & {38} & {55} & {99} & {75} & {175}\\
{16} & {13} & {36} & {39} & {56} & {100} & {76} & {177}\\
{17} & {14} & {37} & {47} & {57} & {111} & {77} & {183}\\
{18} & {12} & {38} & {49} & {58} & {113} & {78} & {187}\\
{19} & {16} & {39} & {50} & {59} & {115} & {79} & {189}\\
{20} & {17} & {40} & {51} & {60} & {117} & {80} & {191}\\
\hline \hline
\end{tabular}
\end{center}
\begin{flushleft}
LG: Layer Group; SG: Space Group.
\end{flushleft}
\end{table}

\section{Example of Lattice-enforced Semimetals
\label{app:leSM}}
Here, we provide details on the leSM example discussed in the main text. We consider a TR-symmetric system in SG 219 with significant spin-orbit coupling. We will establish that for a particular lattice specification, a semimetallic behavior is unavoidable at a filling $\nu=4$ although band insulators are generally possible at this filling for the present symmetry setting \cite{PNAS, PRL}. This arises from the fact that, given the available symmetry irreps specified by the lattice, corresponding to an element $\vec A \in \{ {\rm AI}\}$, there is no way to satisfy all the compatibility relations at the filling $\nu=4$, i.e.~$\vec A \neq \vec B_v + \vec B_c$ for any non-zero $\vec B_v,\vec B_c \in \{ {\rm BS}\}$ satisfying the physical condition of non-negativity.

We consider a lattice in Wyckoff position $a$, which contains two sites at $\vec r_1 \equiv \{0,0,0\}$ and $\vec r_2 \equiv \{ 1/2,0,0\}$ in the unit cell. The two sites are related by a glide symmetry, and the site-symmetry group for each site is given by the point group $T$ (i.e., the orientation-preserving symmetries of a tetrahedron, also known as the chiral tetrahedral symmetry group). We suppose the physically relevant degrees of freedom arise from the three $p_{x,y,z}$ orbitals on each site, which together with electron spin leads to a 6-dimensional local Hilbert space. We will let $\vec L$ and $\vec S$ respectively denote the orbital and spin angular momentum operators in the single-particle basis.

As described in the main text, we consider a TR-symmetric system with a strong crystal-field splitting:
\begin{equation}\begin{split}\label{eq:}
H_{\Delta} = \Delta \sum_{\vec r: 
\text{all sites}} \vec c^\dagger_{\vec r} \left(\vec L \cdot \vec S \right) \vec c_{\vec r},
\end{split}\end{equation}
where $\vec c_{\vec r}$ represent the $6$-dimensional (row) vector corresponding to the internal degrees of freedom. One can verify that when $\Delta>0$, $H_{\Delta}$ splits the local energy levels to a total spin-1/2 doublet lying below the total spin-3/2 multiplet.
{
While we have chosen $H_{\Delta}$ to conserve the total spin $\vec L + \vec S$ for convenience, such conservation is not required by the local symmetry, which is described by the point group $T< {\rm SO}(3)$. Therefore, the total spin quantum numbers are not a priori good quantum numbers for the problem at hand. However, one can verify that the spinful, time-reversal symmetric irreps of $T$ coincide with the total spin decomposition described above \cite{Bradley}, and hence insofar as symmetries are concerned $H_{\Delta}$ is a sufficiently generic crystal-field Hamiltonian.
We also note that, if time-reversal symmetry is broken, the four-fold degenerate states originating from the total spin-3/2 states can be further split.
}

As discussed in the main text, we are interested in the system arising from half filling the four-fold degenerate local energy levels. { To this end, we assume $\Delta$ is the dominant energy scale in the problem, which implies the low-lying doubly degenerate states can be decoupled from the description of the system as long as they are fully-filled. This leaves behind the four-fold degenerate energy levels, which we assume are half-filled. As there are two symmetry-related sites in each unit cell, these considerations altogether imply that the band structure around the Fermi-energy is described by an effective eight-band tight-binding model at filling $\nu=4$.}

Next we consider a nearest-neighbor hopping term 
\begin{equation}\begin{split}\label{eq:}
H_{t,\lambda} =  \sum_{g\in \mathcal G} g \left (\vec c^\dagger_{\vec r_1} \left(t + \lambda\, \vechat x \cdot (\vec L \times \vec S) \right) \vec c_{\vec r_2} \right)g^\dagger+ {\rm h.c.},
\end{split}\end{equation}
where ${\rm h.c.}$ denotes Hermitian conjugate, and the notation $\sum_{g \in \mathcal G} g \left( \dots \right)g^\dagger $ denotes all the terms generated by transforming the bond in the parenthesis by the symmetry elements of the SG $\mathcal G$.

The band structure of the full Hamiltonian $H = H_{\Delta} + H_{t,\lambda}$ is shown in Fig.~\ref{fig:Applications}e, with parameters $(t/\Delta, \lambda/\Delta) = (0.01,0.05)$. 
{
Note that we have only shown the eight-bands near the Fermi energy; four fully filled bands arising from the doubly degenerate local orbitals are separated in energy by $\mathcal O(\Delta)$.}
As our computation dictates, the lattice specification gives rise to energy bands that are necessarily gapless along the high-symmetry lines at filling $\nu=4$. Interestingly, note that the lattice-enforced gaplessness is of a more subtle flavor: Unlike spinless graphene, where the the gaplessness is enforced by the dimensions of the irreps involved, here all the irreps have dimensions $\leq 4$, and therefore the impossibility of finding a BS at $\nu=4$ is reflected in the connectivity of the energy bands.

In closing, we remark that the notion of leSM is not as robust as the other notions we introduced in this work, say feSM or  reQBI. Specifically, the (semi-)metallic behavior of the system is protected by the assumed microscopic degrees of freedom, which is only sensible assuming a certain knowledge about the energetics of the problem. Under stacking of a trivial phase, say when we incorporate into the description a set of fully filled bands corresponding to an atomic insulator, the enforced gaplessness may become unstable as these apparently inert degrees of freedom can also supply the representations needed to open a gap at the targeted filling.
This can be readily seen from the example above: If we switch the sign of $\Delta$, the same electron filling will now correspond to the full filling of the four-fold degenerate multiplet on each site, which leads to an atomic insulator. Such instability should be contrasted with, say, the notion of reQBIs, which by definition remains nontrivial as long as the extra degrees of freedom we introduce are in the trivial class, i.e.~correspond to AIs.\\

\section{Filling-enforced Quantum Band Insulators
\label{app:feSM}}

In this note we provides details on the feQBIs found in this work, which we briefly mentioned in the main text.  
A band insulator, invariant under a set of symmetries (including an SG $\mathcal{G}$), is called a feQBI if the number of the occupied bands (i.e., the filling) is different from that of any AIs with the same symmetry.  Thus feQBIs are a special case of reQBIs.  As all possible TR-symmetric feQBIs have been discussed in Ref.~\cite{SA}, we will mainly consider systems without TR invariance.  

\subsection{TR-breaking spinless filling-enforced quantum band insulators}
Let us start with feQBIs in the system of spinless fermions, which necessarily break the TR symmetry.  This is because of the following reason: If there existed a TR-symmetric feQBI for spinless fermions, we could immediately construct a TR-symmetric feQBI for spinful fermions with the spin SU(2) symmetry.  However, we know that the latter does not exist according to Ref.~\cite{SA}.  

As a strategy to find feQBIs, one can focus on the electron fillings among elements of $\{ {\rm BS}\}$ and $\{ {\rm AI}\}$.  Whenever there is a mismatch between them, it indicates that certain nontrivial element of $X_{\rm BS}$ can be diagnosed simply using the electron filling. 
This question can again be tackled efficiently using the vector-space like structure of $\{ {\rm BS} \}$ and $\{ {\rm AI}\}$.  We indeed found a mismatch for several SGs for spinless fermions without TR symmetry.  

The mismatch can possibly occur only in those 12 SGs listed in Table~\ref{feQBI11}. To see this, we should focus on (the maximal) fixed-point-free subgroups $\Gamma$ of $\mathcal{G}$ (not to be confused with the $\Gamma$ point in BZ).
For a fixed-point-free SG $\Gamma$, it is easy to show that a certain number of bands must cross with each other somewhere at high-symmetry points or on high-symmetry lines, and that the BI fillings are integer multiples of $\nu_\Gamma>1$.  Hence, if $\mathcal{G}$ contains $\Gamma$ as a subgroup, we know that any $\mathcal{G}$-symmetric BI should have a filling $\nu_\Gamma n$.  For all 218 SGs not listed in Table~\ref{feQBI11}, there exists a $\mathcal{G}$-symmetric AI with the filling $\nu_\Gamma$.  Hence, there cannot be any mismatch of the filling between $\{ {\rm BS}\}$ and $\{ {\rm AI}\}$ for them.  On the other hand, for 8 SGs out of the 12 SGs in Table~\ref{feQBI11}, we found that the filling of the elements of $\{ {\rm BS}\}$ is $2\mathbb{Z}$, while that of $\{ {\rm AI}\}$ is $4\mathbb{Z}$.  Hence we can expect a feQBI at filling $\nu=4n+2$.  

The case of SG {220} is more nontrivial.  In Table~\ref{feQBI11}, we show the set of fillings $\mathcal{S}_{\mathcal{G}}^{\text{AI}}$ ($\mathcal{S}_{\mathcal{G}}^{\text{BI}}$) that correspond to at least one AI (BI).  This set was computed by the following way.  Let $\nu_{\vec{x},r}$ be the filling of the AI arising from fully occupying a site-symmetry group irrep $u_{\vec{x}}^r$ of the position $\vec{x}$.  Then we take superpositions with non-negative integer coefficients: $\mathcal{S}_{\mathcal{G}}^{\text{AI}}=\{\sum_{\vec{x},r}m_{\vec{x},r}\nu_{\vec{x},r}~:~m_{\vec{x},r}\in\mathbb{Z}_{\geq0}\}$, which is, in principle, different from the fillings for $\{ {\rm AI} \} \cap \mathbb Z^{D}_{\geq 0}$.  On the other hand, $\mathcal{S}_{\mathcal{G}}^{\text{BI}}$ is simply the fillings for $\{ {\rm BI} \} \cap \mathbb Z^{D}_{\geq 0}$.  In the case of  $\mathcal{G}=\text{{220}}$, although the filling for $\{ {\rm AI}\}$ and $\{ {\rm BS}\}$ are both $2\mathbb{Z}$, $\mathcal{S}_{\mathcal{G}}^{\text{AI}}$ and $\mathcal{S}_{\mathcal{G}}^{\text{BI}}$ do not agree with each other and there is a feQBI at filling $\nu=4$:
\begin{eqnarray}
\mathcal{S}_{\mathcal{G}}^{\text{AI}}&=&2\mathbb{Z}_{\geq 0}\setminus\{2,4,10\}=\{0,6,8,12,14,16,\ldots\},\\
\mathcal{S}_{\mathcal{G}}^{\text{BI}}&=&2\mathbb{Z}_{\geq 0}\setminus\{2\}=\{0,4,6,8,10,12,14,16,\ldots\}.
\end{eqnarray}

Table~\ref{feQBI12} summarizes TR breaking spinless feQBIs in $8+1=9$ SGs identified above.  Note that these new feQBI examples are, in a sense, more intriguing than the ones we discussed in Ref.~\cite{SA}.  In our previous examples assuming TR-symmetric spinful electrons, the filling condition was enriched by Kramers degeneracy, and feQBIs were discovered at odd-integer site fillings.
In these new examples we discovered, however, the site filling is fractional for any choice of SG-symmetric lattices. 
Superficially, this might appear to be at odds with the conventional wisdom that gapped phases at fractional fillings are associated with either discrete symmetry breaking or intrinsic topological order--both only possible in the presence of interactions.  Rather, these examples highlight the fact that spatial symmetries can lead to intrigue constraints between filling and phases, as was discussed in  Ref.~\cite{PNAS}, which, in fact, correctly predicted that symmetry-protected topological phases might be possible for these systems at such fractional fillings.


\begin{table*}
\caption{\label{feQBI11} \bf Band insulator fillings for some special space groups}
\begin{center}
\begin{tabular}{c|ccc}\hline\hline
Space group $\mathcal{G}$	&$\mathcal{S}_{\mathcal{G}}^{\text{AI}}$& $\mathcal{S}_{\mathcal{G}}^{\text{BI}}$ &$\mathcal{S}_{\mathcal{G}}^{\Gamma}$   \\
\hline
{106},{110}     & $4\mathbb{Z}_{\geq 0}$ & $2\mathbb{Z}_{\geq 0}$ & $2\mathbb{Z}_{\geq 0}$ \\
{73},{133},{142}, {206}, {228}     & $4\mathbb{Z}_{\geq 0}$ & $2\mathbb{Z}_{\geq 0}\setminus\{2\}$ & $2\mathbb{Z}_{\geq 0}$ \\
{135}     & $4\mathbb{Z}_{\geq 0}$ & $4\mathbb{Z}_{\geq 0}$ & $2\mathbb{Z}_{\geq 0}$  \\\hline
{199}, {214}   & $2\mathbb{Z}_{\geq 0}\setminus\{2\}$ & $2\mathbb{Z}_{\geq 0}\setminus\{2\}$ & $2\mathbb{Z}_{\geq 0}$  \\
{220}   & $2\mathbb{Z}_{\geq 0}\setminus\{2,4,10\}$ & $2\mathbb{Z}_{\geq 0}\setminus\{2\}$ & $2\mathbb{Z}_{\geq 0}$ \\
{230}   & $4\mathbb{Z}_{\geq 0}\setminus\{4\}$ & $2\mathbb{Z}_{\geq 0}\setminus\{2,4\}$ & $2\mathbb{Z}_{\geq 0}$ \\
\hline\hline
\end{tabular}
\end{center}
\begin{flushleft}
AI: Atomic Insulator; BI: Band Insulator;
$\mathcal{S}_{\mathcal{G}}^{\text{AI}}$: the set of fillings for AI;
$\mathcal{S}_{\mathcal{G}}^{\text{BI}}$: the set of fillings for BI;
$\mathcal{S}_{\mathcal{G}}^{\Gamma}$: is the set of BI fillings for fixed-point free subgroup $\Gamma$ of $\mathcal{G}$.\\
Note--By definition, we have $\mathcal{S}_{\mathcal{G}}^{\text{AI}}\leq \mathcal{S}_{\mathcal{G}}^{\text{BI}}\leq  \mathcal{S}_{\mathcal{G}}^{\Gamma}$, and in fact $\mathcal{S}_{\mathcal{G}}^{\text{AI}}=\mathcal{S}_{\mathcal{G}}^{\text{BI}}=\mathcal{S}_{\mathcal{G}}^{\Gamma}$ for all 218 SGs not listed here.  
\end{flushleft}
\end{table*}

\begin{table}
\caption{\label{feQBI12} {\bf Summary of results for the 12 space groups listed in Table~\ref{feQBI11}}.
}
\begin{center}
\begin{tabular}{c|cc}\hline\hline
Space group $\mathcal{G}$	& $X_{\rm BS}$ & Generator of $X_{\rm BS}$  \\
\hline
{106},{110}     &  $\mathbb{Z}_2$ & feQBI at filling $2$ \\
{73},{133},{142}, {206}, {228}      & $\mathbb{Z}_2$ & feQBI at filling $6$\\
{135}      & $1$ & --- \\\hline
{199}, {214}   & $1$ & --- \\
{220}   &  $\mathbb{Z}_2$ & feQBI at filling $4$\\
{230}  &  $\mathbb{Z}_2$ & feQBI at filling $6$, $10$\\
\hline\hline
\end{tabular}
\end{center}
\begin{flushleft}
$X_{\rm BS}$: the quotient group between the group of band structures and atomic insulators.
\end{flushleft}
\end{table}

\subsection{A tight-binding model}
As we have cautioned before, a nontrivial BS can generally be either a reSM or a reQBI. To establish that the filling mismatch indeed leads to feQBIs, one has to further assert that a band gap is possible at all generic momenta in the BZ.  To achieve this goal we explicitly constructed a tight-binding model for each of feQBIs in Table~\ref{feQBI12}.  As a concrete example, let us discuss the case of SG {106}, which hosts feQBI at filling $\nu = 2$. 

Consider a tight-binding model (4 band model) defined on $\mathcal{W}^{\text{{106}}}_b$ with one orbital per site.  Here, $\mathcal{W}^{\text{{106}}}_b$ means the Wyckoff position with the Wyckoff letter $b$ in  Ref.~\cite{ITC}:
\begin{equation}
\textstyle(0,\frac{1}{2},z), (\frac{1}{2},0,z+\frac{1}{2}), (\frac{1}{2},0,z), (0,\frac{1}{2},z+\frac{1}{2})
\end{equation}
The site symmetry of this Wyckoff position is the $\pi$-rotation about the $z$-axis.  We use a $p$ orbital that flips sign under the rotation.  These informations specify the representation $[U_{\vec{k}}(g)]_{\sigma'j,\sigma i}\equiv [U_{\vec{x}}^r(g)]_{\sigma'j\vec{k}',\sigma i\vec{k}}$ of $\mathcal{G}=\text{{106}}$ in the tight-binding model as discussed in \ref{app:AI}. One can construct $U_{\vec{k}}(g)$ using  Eq.~\eqref{eq:induced2}.  (In this particular example, $i,j$ label can be dropped since the orbital is a 1D representation). In this notation, $U_{\vec{k}}(g)$ satisfies multiplication rule $U_{p_{g_2}\vec{k}}(g_1)U_{\vec{k}}(g_2)=U_{\vec{k}}(g_1g_2)$.

We construct a $\mathcal{G}$-symmetric Hamiltonian $H_{\vec{k}}$ in two steps: First, we choose an arbitrary $4$ by $4$ Hermitian matrix $h_{\vec{k}}$; next, we symmetrize $h_{\vec{k}}$  by performing the summation:
\begin{equation}
H_{\vec{k}}=\sum_{g\in\mathcal{G}/T}U_{\vec{k}}(g)^\dagger h_{p_g\vec{k}}U_{\vec{k}}(g).
\end{equation}
After the summation, $H_{\vec{k}}$ automatically fulfills the symmetry requirement, i.e., $H_{p_g\vec{k}}U_{\vec{k}}(g)=U_{\vec{k}}(g)H_{\vec{k}}$.

\begin{figure}[h]
\begin{center}
{\includegraphics[width=0.45 \textwidth]{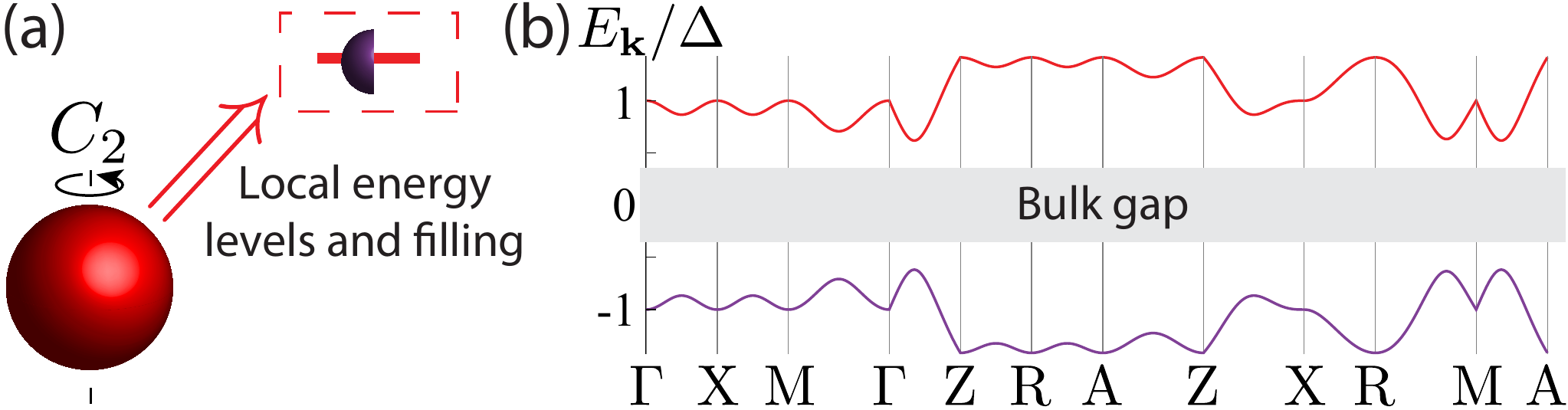}} 
\caption{
\label{fig:feQBI}
{\bf Filling-enforced quantum band insulators in unconventional symmetry settings.}
Here, we focus on one such example found among spinless systems in the tetragonal space group $106$ without time-reversal symmetry. (a) For this space group, a maximal-symmetry site (red sphere) is invariant under a $C_2$ rotation, and symmetries require that there are at least four such sites in the unit cell. We consider a filling of $\nu = 2$, corresponding to a site filling of $1/2$. Note that, given the space group symmetries, any other choice of lattices correspond to site fillings $\leq 1/2$.
(b) Nonetheless, a band insulator is possible at such filling, as shown in the plotted band structure. Each band shown is doubly-degenerate, but such degeneracy originates from nonessential additional symmetries in the simple tight-binding model we constructed, and therefore can be lifted.
}
\end{center}
\end{figure}

To realize an example of feQBI for {106}, the following choice of $h_{\vec{k}}$ works
\begin{eqnarray}
h_{\vec{k}}&=&
\Delta
\begin{pmatrix}
(\cos k_x - \cos k_y)&1&0&-i\\
1&0&0&0\\
0&0&0&-i\\
i&0&i&0
\end{pmatrix},
\end{eqnarray}
where $\Delta $ sets the energy scale of the problem.
The symmetrized Hamiltonian $H_{\vec{k}}$ has a very large gap.  At each $\vec{k}$, let $\delta E_{\vec{k}}$ be the band gap between the second and the third band and $W_{\vec{k}}$ be the band width (the energy difference between the lowest and the highest band). Then we found $\delta E=0.33W$ where $\delta E=\min_{\vec{k}}\delta E_{\vec{k}}$ and $W=\max_{\vec{k}}W_{\vec{k}}$.   An example band structure is plotted in Fig.~\ref{fig:feQBI}, which demonstrates that even at a filling of half an electron per site, a band gap is nonetheless possible.  We also found that the Chern numbers of the tight-binding model is zero.   

In Table~\ref{feQBI2}, we list some possible symmetry settings for feQBIs in other SGs.  
\begin{table*}
\caption{
\bf Possible symmetry setting for the filling-enforced quantum band insulators in Table~\ref{feQBI12}.
\label{feQBI2}}
\begin{center}
\begin{tabular}{cccccccc}\hline\hline
Space group & Wyckoff position  & orbital& $m_{\text{tot}}$ & $\nu$  \\\hline
{106}&$b$&$p^*$&4&2 \\\hline
{110}&$a$&$p^*$&4&2\\\hline
{133}&$a+e$&$s$&12&6\\\hline
{220}&$c$&$s$&8&4\\\hline
{228}($>${142}, {73})&$d$&$s$&12&6\\\hline
{230}($>$ {206})&$a$&$s+p^{**}$&16&6,10\\
\hline\hline
\end{tabular}
\end{center}
\begin{flushleft}
feQBI: filling-enforced quantum band insulators;
$m_{\text{tot}}$: the total number of bands in this setting;
$\nu$: electron fillings per primitive unit cell for which feQBIs are possible;\\
$^*$ The site symmetry group for the Wyckoff position $b$ of {106} and the Wyckoff position $a$ of {110} are an order $2$ group, and the $p$ orbital refers to the representation with $-1$ for the non-identity element.\\
$^{**}$ The site symmetry group for the Wyckoff position $a$ of {230} is $\bar{3}$ group, generated by the improper three-fold rotation $IC_3$. The $p$ orbital refers to the one that has $e^{i\frac{2\pi}{6}}$ eigenvalue of $IC_3$.
\end{flushleft}
\end{table*}

\subsection{TR-symmetric filling-enforced quantum band insulators}
Finally, we briefly comment on the implications of the present work on the feQBIs we introduced in  Ref.~\cite{SA}, which arise in systems with TR symmetry and significant spin-orbit coupling.
There, the focus of study are SGs {199}, {214}, {220} and {230}, which have an intriguing property in the ratios between the Wyckoff position multiplicities: They were dubbed Wyckoff-mismatched as some Wyckoff multiplicities are not integer multiples of the smallest one. In these systems, it was realized that QBIs are possible at non-atomic fillings, and hence the name feQBIs.

However, in the analysis of Ref.~\cite{SA} the mismatch between BSs and AIs is discussed in terms of their corresponding physical fillings, which as we have explained is a stronger condition than that exposed using only the abelian group structures of $\{ {\rm BS} \}$ and $\{ {\rm AI} \}$. As a result, these feQBIs could be trivial in $X_{\rm BS}$. This is indeed the case for some of them: both {199} and {214} have no nontrivial BSs in $X_{\rm BS} =\mathbb Z_1$. (By our convention, they are omitted from Table \ref{tab:Spinful_TRX} of the main text.)

It remains to study {220} and {230}, which have $X_{\rm BS} = \mathbb Z_2$ and $\mathbb Z_4$ respectively. We found that for both cases, some feQBIs are again in the trivial class, i.e.~they can be understood as integer combinations of AIs (but with negative coefficients, resulting in unrealizable AI upon enforcing the physical conditions). However, the nontrivial class for {220} can be represented by a feQBI filling $\nu=20$. More interestingly, the entry $2\in \mathbb Z_4$ for {230} can be represented by a feQBI at filling $\nu=8$. Since {230} is centrosymmetric, as we have argued in the main text the generator $1\in \mathbb Z_4$ has to be identified with the strong TI. This implies some $\nu = 8$ feQBIs in {230} realize the doubled strong TI phase we discussed in the main text, which has inversion-protected nontrivial entanglement signature, albeit no physical surface state is expected. 

All in all, we found that generally the TR-symmetric feQBIs do not have any simple relationship with $X_{\rm BS}$, although there are indeed examples which are nontrivial from both perspectives.

\section{Representation-enforced Quantum Band Insulators and Semimetals
\label{app:X_Relate}}
\subsection{General relation}
As we have discussed in the main text, our notion of a BS is compatible with both band insulators and semimetals, as long as a continuous band gap is sustained at all high-symmetry momenta. In particular, the nontrivial entries in $X_{\rm BS}$ can sometimes correspond to reSM, which are guaranteed to be semimetallic due to the specification of the symmetry content. These systems are exemplified by the 3D systems with inversion but not TR symmetries \cite{Ari, Bernevig}.

Since reSMs are, by definition, also diagnosable using the representation content, one can systematically isolate them from $\{ {\rm BS}\}$. However, it is important to realize that reSMs do not form a subgroup of $\{ {\rm BS}\}$, since stacking two of them (say) may lead to a band insulator \cite{Ari,Bernevig}. In contrast, it is guaranteed that stacking two band insulators will lead to yet another band insulator. This suggests that we should identify the subgroup $\{ {\rm BI} \} \leq \{{\rm BS}\}$. Further observe $\{{\rm AI}\} \leq \{ {\rm BI}\}$, it is natural to define the following further diagnosis of the elements in $\{ {\rm BS}\}$:
\begin{equation}\begin{split}\label{eq:}
X_{\rm SM} \equiv  \frac{\{{\rm BS} \} }{\{{\rm BI} \} };~~~~~
X_{\rm BI} \equiv  \frac{\{{\rm BI} \} }{\{{\rm AI} \} }.
\end{split}\end{equation}
The nontrivial entries in $X_{\rm SM}$ and  $X_{\rm BI}$  respectively correspond to reSMs and reQBIs.
As we have alluded to, the nontrivial elements in $X_{\rm BS}$ are either reSMs or reQBIs, and hence, unsurprisingly, the three objects $X_{\rm BS}$, $X_{\rm SM}$ and $X_{\rm BI}$ are not independent. From definitions, one can check that
\begin{equation}\begin{split}\label{eq:}
X_{\rm SM} = \frac{X_{\rm BS} }{X_{\rm BI}}.
\end{split}\end{equation}
Or in a more formal language, $X_{\rm BS}$ can be viewed as a central extension of $X_{\rm SM}$ by $X_{\rm BI}$. Curiously, this extension is generally nontrivial. As a concrete example, consider the $\mathbb Z_4$ factor in $X_{\rm BS} = (\mathbb Z_2)^3 \times \mathbb Z_4 $ for SG {2} (inversion only) assuming no TR symmetry. From Refs.~\cite{Ari, Bernevig}, we see that the generator of this factor is a reSM (with inversion related Weyl points), and the twice of that, corresponding to $2\in \mathbb Z_4$, is a reQBI with a quantized magnetoelectric response of $\theta =\pi$. This shows that for this particular example, 
\begin{equation}\begin{split}\label{eq:}
X_{\rm BI} = (\mathbb Z_2)^4;~~~X_{\rm SM} = \mathbb Z_2,
\end{split}\end{equation}
and the $\mathbb Z_4$ factor in  $X_{\rm BS}$ originates from the nontrivial extension of $\mathbb Z_2$ by $\mathbb Z_2$.

\subsection{Representation-enforced quantum band insulators}
As we will discuss in the following, given any symmetry setting one can systematically study all reSMs using symmetry arguments together with knowledge on the generic stability of Femri surfaces. For instance, for centrosymmetric systems with TR symmetry and significant spin-orbit coupling, stable band degeneracy must happen at a high-symmetry momentum, and therefore we can rule out the possibility of reSMs in these problems, i.e.~for such settings we have $X_{\rm SM} = \mathbb Z_1$ and hence $X_{\rm BI} =X_{\rm BS}$. However, even after $X_{\rm BI}$ is obtained our approach is still based on symmetry-labels in nature, and therefore does not necessarily detect all nontrivial phases. An important future direction is to incorporate the tenfold way classification into our current framework \cite{FreedMoore}, akin to the arguments given in Ref.~\cite{Combinatorics}. 

\subsection{Representation-enforced semimetals}
For a tight-binding model with inversion but not TR symmetry, some combinations of the parity eigenvalues at TRIMs predict the existence of Weyl points somewhere in the interior of the BZ \cite{Ari,Bernevig}.  
The semimetallic behavior of such systems is enforced by the specification of the representations, and we refer to such systems as reSMs. 
In this section, we ask whether similar phenomena occur for other SG, i.e., given an SG and a set of integers $\vec{n}$ for irreps at high-symmetry momenta, we ask if there are enforced gap closing somewhere at non-high-symmetry points in the BZ.
For simplicity, in the following we answer this question for three-dimensional system without TR symmetry, such that the topologically-protected gap closing at non-high-symmetry momenta corresponds to Weyl points.

Let us consider a unit sphere $S^2$ around the $\Gamma$ point of the BZ.  The radius of the sphere is set to be much smaller than any reciprocal lattice vectors.  If there are Weyl points at  generic momenta in the BZ (i.e.~not high-symmetry), we should be able to move them to the surface of this sphere without breaking any symmetry or changing the value of $\vec{n}$. Note that, in the following `sphere' always refer to $S^2$, which does not include the interior of the sphere.

An SG element $g\in\mathcal{G}$ moves a point $\vec{k}$ on $S^2$ to $p_g\vec{k}$ on $S^2$.  For this transformation, the translation part $\vec{t}_g$ of $g$ does not show up at all and hence what is really important is the point group $\mathcal{P}\simeq\mathcal{G}/T$ of the SG, whose elements are the orthogonal matrices $p_g$ ($g\in\mathcal{G}$).  There are only 32 crystallographic point groups in 3D, and we will systematically discuss them in the following.  One can introduce the notion of the little group, the symmetry orbit, and Wyckoff positions to this $S^2$ in the same way as before. 
Given $\vec{k}$ on $S^2$, the little group $\mathcal{G}_{\vec k}$ is defined as
\begin{equation}
\mathcal{G}_{\vec k}=\{p_g\in \mathcal{P}~:~p_g\vec{k}=\vec{k}\}.
\end{equation}
The symmetry orbit of $\vec{k}$ (aka the star of $\vec{k}$) is defined as $\{p_g\vec{k}~:~p_g\in\mathcal{P}\}$.  Also, two points $\vec{k}_1$ and $\vec{k}_2$ belong to the same (momentum-space) Wyckoff position iff there exists $g\in\mathcal{G}$ such that $\mathcal{G}_{\vec{k}_2}=p_g\mathcal{G}_{\vec{k}_1}p_g^{-1}$.  
Finally, let us define the irreducible part (or the fundamental domain) $F$ of the sphere. The fundamental domain $F$ tessellates the $S^2$ under the action of $\mathcal{P}$. Namely, any point on the sphere can be uniquely represented as $p_g\vec{k}$, where $\vec{k}\in F$ and $p_g\in \mathcal{P}$.  

With this preparation, let us discuss the stability of the Weyl points against symmetry preserving deformations.  Suppose that the fundamental domain $F$ contains a Weyl point with the chirality $+1$. 
(By assumption, generically this Weyl point does not sit at the boundary of $F$.) Then there must be at least $|\mathcal{P}|$ Weyl points on the sphere in total.  The domain $p_gF$ will contain a Weyl point with the chirality $\text{det}[p_g]$.  Suppose first that $\text{det}[p_g]=+1$ for all $p_g\in\mathcal{P}$. This is the case for 11 out of 32 point groups in 3D.  In this case, since the net chirality in the entire BZ must vanish, there must be another Weyl point in $F$ with the chirality $-1$ (that belongs to a separate symmetry orbit), and the two Weyl points with the opposite chirality will freely annihilate with each other. This implies reSMs cannot exist for SGs with these point groups.

Next we discuss the remaining $32-11 = 21$ point groups, which always have half of the elements having $\text{det}[p_g]=-1$. Again the number of symmetry-related Weyl points is $|\mathcal{P}|$, but now $|\mathcal{P}|/2$ of them has $+1$ chirality and the other half has the $-1$ chirality.  
A prioi, this could be compatible with a reSM, since the net Weyl charge in the entire BZ vanishes. However, this still requires further symmetry analysis.
We ask if one can move Weyl points on $S^2$ and merge pairs of them with  opposite chirality without breaking the SG symmetry or changing the number of irreps $\vec{n}$. We found that $18$ of the $21$ remaining point groups contain at least one mirror reflection symmetry and there exists a Wyckoff position on the sphere which is symmetric only under the mirror.  For these point groups, two Weyl points with the opposite chirality should be able to annihilate at this Wyckoff position. Provided this is true, this again rules out reSMs in these settings. In the following, we will first assume the validity of this claim and study the remaining point groups; a more thorough analysis is left for future works.

The remaining three point groups are $\bar{1}$, $\bar{3}$, and $\bar{4}$ in the Hermann-Mauguin notation.  Among them, $\bar{1}$ is the point group generated by the inversion symmetry, and we know that Weyl points can be protected by $\vec{n}$ in this case \cite{Ari,Bernevig}.  $\bar{3}$ contains $\bar{1}$ as a subgroup, and therefore also contains the rotation $C_3$. Due to the relation between the $C_3$ eigenvalues and the Chern number (for any momentum plane perpendicular to the rotation axis) \cite{Chern_Rotation}, one can show that the only reSMs for $\bar{3}$ are those that are diagnosed by the $\bar{1}$ subgroup.
As such, the only possibly nontrivial candidate is thus the point group $\bar{4}$, which is the point group of the SG symmetry {81} and {82}.  We looked at {81} for the spinful but TR breaking setting, and indeed found an reSM.  $X_{\rm BS}$ for {81} is $(\mathbb{Z}_2)^2\times\mathbb{Z}_4$. The reSM is the generator of one of the two $\mathbb{Z}_2$ factor. In fact, this reSM  in SG {81} can also be understood from the Chern number arguments in Ref.~\cite{Chern_Rotation}.

In closing, we remark that the analysis of reSM has to be modified in other symmetry setting, since depending on symmetries topologically stable gaplessness may not be of the Weyl type. For instance, nodal lines are stable in time-reversal and inversion symmetric systems with negligible spin-orbit coupling, and therefore the argument above has to be modified from the motion of points to nodal rings. In addition, similar to the point-group symmetries we discussed, TR will also constrain the possible distribution of the gap closing, and therefore have to be taken into account in the analysis of reSMs in TR-symmetric system.

\section{Supplementary tables
\label{app:SuppTab}}
Here, we provide tables for the computed values of $d\equiv d_{\rm BS} = d_{\rm AI}$ and $X_{\rm BS}$ for 3D systems without TR symmetries, and for quasi-2D and 1D systems described respectively by layer group and rod group symmetries.

\onecolumngrid
\input{tab_SpinfulNoTRd}
\input{tab_SpinlessNoTRd}

\input{tab_SpinfulNoTRX}
\input{tab_SpinlessNoTRX}

\input{tab_LG_SpinfulTRd}
\input{tab_LG_SpinlessTRd}
\input{tab_LG_SpinfulNoTRd}
\input{tab_LG_SpinlessNoTRd}

\input{tab_RG_SpinfulTRd}
\input{tab_RG_SpinlessTRd}
\input{tab_RG_SpinfulNoTRd}
\input{tab_RG_SpinlessNoTRd}

\input{tab_LG_SpinfulTRX}
\input{tab_LG_SpinlessTRX}
\input{tab_LG_SpinfulNoTRX}
\input{tab_LG_SpinlessNoTRX}

\twocolumngrid
\clearpage
\bibliography{references}

\end{document}

%% file: tab_SpinfulNoTRd.tex
\begin{center}
\begin{table}[h]
\caption{
\bf Characterization of band structures for systems with significant spin-orbit coupling and no time-reversal symmetry.}
\begin{tabular}{c|c} \hline \hline
$d$ & Space groups \\
\hline
$1$  &  1, 4, 7, 9, 16, 19, 22, 23, 25, 26, 27, 29, 33, 36, 38, 39, 42, 44, 45, 76, 78, 93, 101, 105, 109, 110\\
     &  144, 145, 169, 170, 180, 181\\
\hline
$2$  &  8, 21, 31, 35, 37, 41, 43, 46, 80, 92, 94, 96, 97, 98, 102, 106, 107, 108, 161, 208, 214\\
\hline
$3$  &  5, 6, 18, 20, 30, 32, 34, 40, 48, 50, 56, 59, 61, 62, 68, 70, 73, 89, 99, 103, 146, 151, 152, 153, 154, 160\\
     &  171, 172, 178, 179, 185, 186, 195, 196, 197, 198, 209, 210, 211, 212, 213\\
\hline
$4$  &  24, 28, 54, 57, 60, 72, 77, 90, 91, 95, 100, 104, 133, 137, 142, 155, 158, 159, 177, 183, 184, 199, 207\\
\hline
$5$  &  3, 14, 17, 52, 63, 64, 67, 79, 111, 112, 115, 116, 119, 120, 121, 126, 130, 134, 138, 156, 157, 173, 182\\
\hline
$6$  &  11, 15, 49, 69, 71, 113, 114, 117, 118, 125, 129, 132, 135, 141, 149, 150, 215, 216, 217, 218, 219\\
\hline
$7$  &  13, 51, 55, 66, 74, 122, 131, 136, 143, 167, 220, 228, 230\\
\hline
$8$  &  58, 65, 75, 88, 140, 163, 165, 222, 223, 224\\
\hline
$9$  &  2, 47, 53, 86, 139, 168, 201, 203, 205, 206, 227\\
\hline
$10$  &  12, 187, 189, 193, 194, 202, 204, 226\\
\hline
$11$  &  82, 85, 124, 148, 166, 200, 225, 229\\
\hline
$12$  &  81, 127, 128, 162, 164, 188, 190\\
\hline
$13$  &  84, 123, 147, 192\\
\hline
$14$  &  191, 221\\
\hline
$15$  &  10\\
\hline
$16$  &  87, 176\\
\hline
$21$  &  174\\
\hline
$24$  &  83\\
\hline
$27$  &  175\\
\hline
\hline
\end{tabular}
\begin{flushleft}
$d$: the rank of the abelian group formed by the set of band structures.
\end{flushleft}
\end{table}
\end{center}

%% file: tab_SpinlessNoTRd.tex
\begin{center}
\begin{table}[h]
\caption{
\bf Characterization of band structures for systems of spinless fermions without time-reversal symmetry.}
\begin{tabular}{c|c} \hline \hline
$d$ & Space groups \\
\hline
$1$  &  1, 4, 7, 9, 19, 29, 33, 76, 78, 144, 145, 169, 170\\
\hline
$2$  &  8, 31, 36, 41, 43, 80, 92, 96, 110, 161\\
\hline
$3$  &  5, 6, 18, 20, 26, 30, 32, 34, 40, 45, 46, 61, 106, 109, 146, 151, 152, 153, 154, 160, 171, 172, 178, 179, 198, 212\\
     &  213\\
\hline
$4$  &  24, 28, 37, 39, 60, 62, 77, 91, 95, 102, 155, 158, 159, 185, 186, 199, 210\\
\hline
$5$  &  3, 14, 17, 27, 42, 44, 52, 56, 57, 79, 94, 98, 101, 104, 108, 156, 157, 173, 182, 196, 197, 214\\
\hline
$6$  &  11, 15, 35, 38, 54, 70, 73, 100, 103, 105, 107, 149, 150, 184\\
\hline
$7$  &  13, 22, 23, 59, 64, 68, 90, 114, 122, 142, 143, 167, 180, 181, 195, 208, 209, 211, 220\\
\hline
$8$  &  21, 58, 63, 75, 88, 97, 113, 130, 137, 163, 165, 183, 219\\
\hline
$9$  &  2, 25, 48, 50, 53, 55, 72, 86, 99, 117, 118, 120, 133, 135, 141, 168, 205, 207, 216, 217, 218, 228, 230\\
\hline
$10$  &  12, 74, 93, 116, 119, 121, 126, 138, 177, 203, 206, 215\\
\hline
$11$  &  66, 82, 85, 148, 166, 201, 222, 227\\
\hline
$12$  &  51, 81, 89, 112, 115, 129, 134, 136, 162, 164, 188, 190\\
\hline
$13$  &  16, 67, 84, 111, 125, 147, 193, 194, 202, 204, 223, 224\\
\hline
$14$  &  49, 128, 226\\
\hline
$15$  &  10, 69, 71, 132, 140, 187, 189\\
\hline
$16$  &  87, 176\\
\hline
$17$  &  124, 192, 200, 225, 229\\
\hline
$18$  &  65, 127, 131, 139\\
\hline
$21$  &  174\\
\hline
$22$  &  221\\
\hline
$24$  &  83, 191\\
\hline
$27$  &  47, 123, 175\\
\hline
\hline
\end{tabular}
\begin{flushleft}
$d$: the rank of the abelian group formed by the set of band structures.
\end{flushleft}
\end{table}
\end{center}

%% file: tab_SpinfulNoTRX.tex
\begin{center}
\begin{table}[h]
\caption{
\bf Symmetry-based indicators of band topology for systems with significant spin-orbit coupling and no time-reversal symmetry.}
\begin{tabular}{c|c} \hline \hline
$X_{\rm BS}$ & Space groups \\
\hline
$\mathbb Z_{2}$  &  3, 11, 14, 48, 49, 50, 52, 53, 54, 56, 57, 58, 59, 60, 61, 62, 63, 64, 66, 67, 68, 70,\\
                 &  72, 73, 74, 77, 79, 111, 112, 113, 114, 115, 116, 117, 118, 119, 120, 121, 122, 125, 126, 129, 130, 133,\\
                 &  134, 137, 138, 141, 142, 162, 163, 164, 165, 166, 167, 171, 172, 201, 203, 205, 206, 215, 216, 217, 218, 219,\\
                 &  220, 222, 224, 227, 228, 230\\
\hline
$\mathbb Z_{3}$  &  143, 173, 188, 190\\
\hline
$\mathbb Z_{4}$  &  69, 71, 75, 124, 128, 132, 135, 136, 140, 202, 204, 223, 226\\
\hline
$\mathbb Z_{6}$  &  168, 192, 193, 194\\
\hline
$\mathbb Z_{8}$  &  139, 225, 229\\
\hline
$\mathbb Z_{2} \times \mathbb Z_{2}$  &  12, 13, 15, 51, 55, 86, 88\\
\hline
$\mathbb Z_{2} \times \mathbb Z_{4}$  &  65, 84, 85, 131, 148, 200\\
\hline
$\mathbb Z_{2} \times \mathbb Z_{12}$  &  147\\
\hline
$\mathbb Z_{3} \times \mathbb Z_{3}$  &  187, 189\\
\hline
$\mathbb Z_{3} \times \mathbb Z_{6}$  &  176\\
\hline
$\mathbb Z_{4} \times \mathbb Z_{4}$  &  87, 127\\
\hline
$\mathbb Z_{4} \times \mathbb Z_{8}$  &  221\\
\hline
$\mathbb Z_{6} \times \mathbb Z_{12}$  &  191\\
\hline
$\mathbb Z_{2} \times \mathbb Z_{2} \times \mathbb Z_{2}$  &  10, 82\\
\hline
$\mathbb Z_{2} \times \mathbb Z_{2} \times \mathbb Z_{4}$  &  81\\
\hline
$\mathbb Z_{2} \times \mathbb Z_{4} \times \mathbb Z_{8}$  &  123\\
\hline
$\mathbb Z_{3} \times \mathbb Z_{3} \times \mathbb Z_{3}$  &  174\\
\hline
$\mathbb Z_{4} \times \mathbb Z_{4} \times \mathbb Z_{4}$  &  83\\
\hline
$\mathbb Z_{6} \times \mathbb Z_{6} \times \mathbb Z_{6}$  &  175\\
\hline
$\mathbb Z_{2} \times \mathbb Z_{2} \times \mathbb Z_{2} \times \mathbb Z_{4}$  &  2, 47\\
\hline
\hline
\end{tabular}
\begin{flushleft}
$X_{\rm BS}$: the quotient group between the group of band structures and atomic insulators.
\end{flushleft}
\end{table}
\end{center}

%% file: tab_SpinlessNoTRX.tex
\begin{center}
\begin{table}[h]
\caption{
\bf Symmetry-based indicators of band topology for systems of spinless fermions without time-reversal symmetry.}
\begin{tabular}{c|c} \hline \hline
$X_{\rm BS}$ & Space groups \\
\hline
$\mathbb Z_{2}$  &  3, 11, 14, 27, 37, 45, 48, 49, 50, 52, 53, 54, 56, 58, 60, 61, 66, 68, 70, 73, 77, 79,\\
                 &  103, 104, 106, 110, 112, 114, 116, 117, 118, 120, 122, 126, 130, 133, 142, 162, 163, 164, 165, 166, 167, 171,\\
                 &  172, 184, 201, 203, 205, 206, 218, 219, 220, 222, 228, 230\\
\hline
$\mathbb Z_{3}$  &  143, 173, 188, 190\\
\hline
$\mathbb Z_{4}$  &  75, 124, 128\\
\hline
$\mathbb Z_{6}$  &  168, 192\\
\hline
$\mathbb Z_{2} \times \mathbb Z_{2}$  &  12, 13, 15, 86, 88\\
\hline
$\mathbb Z_{2} \times \mathbb Z_{4}$  &  84, 85, 148\\
\hline
$\mathbb Z_{2} \times \mathbb Z_{12}$  &  147\\
\hline
$\mathbb Z_{3} \times \mathbb Z_{6}$  &  176\\
\hline
$\mathbb Z_{4} \times \mathbb Z_{4}$  &  87\\
\hline
$\mathbb Z_{2} \times \mathbb Z_{2} \times \mathbb Z_{2}$  &  10, 82\\
\hline
$\mathbb Z_{2} \times \mathbb Z_{2} \times \mathbb Z_{4}$  &  81\\
\hline
$\mathbb Z_{3} \times \mathbb Z_{3} \times \mathbb Z_{3}$  &  174\\
\hline
$\mathbb Z_{4} \times \mathbb Z_{4} \times \mathbb Z_{4}$  &  83\\
\hline
$\mathbb Z_{6} \times \mathbb Z_{6} \times \mathbb Z_{6}$  &  175\\
\hline
$\mathbb Z_{2} \times \mathbb Z_{2} \times \mathbb Z_{2} \times \mathbb Z_{4}$  &  2\\
\hline
\hline
\end{tabular}
\begin{flushleft}
$X_{\rm BS}$: the quotient group between the group of band structures and atomic insulators.
\end{flushleft}
\end{table}
\end{center}

%% file: tab_LG_SpinfulTRd.tex
\begin{center}
\begin{table}[h]
\caption{
\bf Characterization of band structures for quasi-two-dimensional systems with significant spin-orbit coupling and  time-reversal symmetry.}
\begin{tabular}{c|c} \hline \hline
$d$ & Layer groups \\
\hline
$1$  &  1, 3, 4, 5, 8, 9, 10, 11, 12, 13, 19, 20, 21, 22, 23, 24, 25, 26, 27, 28, 29, 30, 31, 32, 33, 34\\
     &  35, 36\\
\hline
$2$  &  39, 43, 45, 46, 54, 56, 58, 60\\
\hline
$3$  &  7, 15, 16, 17, 38, 40, 41, 42, 44, 48, 49, 50, 53, 55, 57, 59, 68, 70\\
\hline
$4$  &  18, 47, 52, 62, 64, 65, 67, 69, 73, 76, 77\\
\hline
$5$  &  2, 6, 14, 37, 63, 79\\
\hline
$6$  &  66, 71, 72\\
\hline
$7$  &  74, 78\\
\hline
$8$  &  51, 61\\
\hline
$9$  &  75, 80\\
\hline
\hline
\end{tabular}
\begin{flushleft}
$d$: the rank of the abelian group formed by the set of band structures.
\end{flushleft}
\end{table}
\end{center}

%% file: tab_LG_SpinlessTRd.tex
\begin{center}
\begin{table}[h]
\caption{
\bf Characterization of band structures for quasi-two-dimensional systems with negligible spin-orbit coupling and time-reversal symmetry.}
\begin{tabular}{c|c} \hline \hline
$d$ & Layer groups \\
\hline
$1$  &  1, 5, 9, 12, 33\\
\hline
$2$  &  4, 10, 13, 29, 32, 34\\
\hline
$3$  &  8, 11, 17, 21, 25, 28, 30, 31, 36\\
\hline
$4$  &  15, 16, 20, 24, 35, 43, 45, 65, 68, 70\\
\hline
$5$  &  2, 3, 7, 54, 56, 58, 60, 67, 69\\
\hline
$6$  &  18, 22, 26, 27, 39, 42, 44, 46, 49, 50, 52, 66, 73\\
\hline
$8$  &  40, 71, 72, 74, 76, 77, 79\\
\hline
$9$  &  14, 19, 23, 38, 41, 48, 53, 55, 57, 59, 62, 64\\
\hline
$10$  &  6, 63, 78\\
\hline
$12$  &  47, 51, 75\\
\hline
$16$  &  80\\
\hline
$18$  &  37, 61\\
\hline
\hline
\end{tabular}
\begin{flushleft}
$d$: the rank of the abelian group formed by the set of band structures.
\end{flushleft}
\end{table}
\end{center}

%% file: tab_LG_SpinfulNoTRd.tex
\begin{center}
\begin{table}[h]
\caption{
\bf Characterization of band structures for quasi-two-dimensional systems with significant spin-orbit coupling and no time-reversal symmetry.}
\begin{tabular}{c|c} \hline \hline
$d$ & Layer groups \\
\hline
$1$  &  1, 5, 9, 12, 19, 23, 27, 28, 29, 30, 33, 35, 36\\
\hline
$2$  &  4, 10, 13, 22, 26, 32, 34, 39, 46\\
\hline
$3$  &  8, 11, 17, 21, 25, 31, 43, 45, 48, 53, 55, 57, 59\\
\hline
$4$  &  15, 16, 20, 24, 38, 41, 54, 56, 58, 60, 62, 64, 76, 77\\
\hline
$5$  &  2, 3, 7, 37, 40, 44, 47, 67, 68, 69, 70\\
\hline
$6$  &  18, 42\\
\hline
$7$  &  65, 78, 79\\
\hline
$8$  &  49, 50, 52, 61, 63, 71, 72\\
\hline
$9$  &  14, 66, 73, 80\\
\hline
$10$  &  6\\
\hline
$14$  &  74\\
\hline
$16$  &  51\\
\hline
$18$  &  75\\
\hline
\hline
\end{tabular}
\begin{flushleft}
$d$: the rank of the abelian group formed by the set of band structures.
\end{flushleft}
\end{table}
\end{center}

%% file: tab_LG_SpinlessNoTRd.tex
\begin{center}
\begin{table}[h]
\caption{
\bf Characterization of band structures for quasi-two-dimensional systems of spinless femrions without time-reversal symmetry.}

\begin{tabular}{c|c} \hline \hline
$d$ & Layer groups \\
\hline
$1$  &  1, 5, 9, 12, 33\\
\hline
$2$  &  4, 10, 13, 29, 32, 34\\
\hline
$3$  &  8, 11, 17, 21, 25, 28, 30, 31, 36\\
\hline
$4$  &  15, 16, 20, 24, 35, 43, 45\\
\hline
$5$  &  2, 3, 7, 67, 68, 69, 70\\
\hline
$6$  &  18, 22, 26, 27, 39, 42, 44, 46, 54, 56, 58, 60\\
\hline
$7$  &  65\\
\hline
$8$  &  40, 49, 50, 52, 71, 72, 76, 77\\
\hline
$9$  &  14, 19, 23, 38, 41, 48, 53, 55, 57, 59, 62, 64, 66, 73\\
\hline
$10$  &  6, 78, 79\\
\hline
$12$  &  47, 63\\
\hline
$14$  &  74\\
\hline
$16$  &  51, 80\\
\hline
$18$  &  37, 61, 75\\
\hline
\hline
\end{tabular}
\begin{flushleft}
$d$: the rank of the abelian group formed by the set of band structures.
\end{flushleft}
\end{table}
\end{center}

%% file: tab_RG_SpinfulTRd.tex
\begin{center}
\begin{table}[h]
\caption{
\bf Characterization of band structures for quasi-one-dimensional systems with significant spin-orbit coupling and  time-reversal symmetry.}
\begin{tabular}{c|c} \hline \hline
$d$ & Rod groups \\
\hline
$1$  &  1, 3, 4, 5, 8, 9, 10, 13, 14, 15, 16, 17, 18, 19, 24, 25, 26, 31, 32, 33, 35, 43, 44, 47, 48, 54\\
     &  55, 57, 58, 63, 64, 66, 67\\
\hline
$2$  &  7, 12, 21, 22, 23, 30, 34, 36, 38, 42, 46, 49, 50, 56, 65, 70\\
\hline
$3$  &  2, 6, 11, 20, 27, 29, 37, 41, 53, 62, 68, 69, 72\\
\hline
$4$  &  40, 52, 59, 71\\
\hline
$5$  &  61, 75\\
\hline
$6$  &  28, 39, 45, 51, 74\\
\hline
$9$  &  60, 73\\
\hline
\hline
\end{tabular}
\begin{flushleft}
$d$: the rank of the abelian group formed by the set of band structures.
\end{flushleft}
\end{table}
\end{center}

%% file: tab_RG_SpinlessTRd.tex
\begin{center}
\begin{table}[h]
\caption{
\bf Characterization of band structures for quasi-one-dimensional systems with negligible spin-orbit coupling and time-reversal symmetry.}
\begin{tabular}{c|c} \hline \hline
$d$ & Rod groups \\
\hline
$1$  &  1, 5, 9, 24, 26, 43, 44, 54, 58\\
\hline
$2$  &  4, 8, 16, 17, 25, 42, 50, 55, 56, 57\\
\hline
$3$  &  2, 3, 7, 10, 12, 14, 19, 23, 31, 33, 35, 36, 47, 48, 49, 63, 67, 70\\
\hline
$4$  &  15, 27, 46, 53, 65, 69\\
\hline
$5$  &  29, 34, 38, 52, 72\\
\hline
$6$  &  6, 11, 13, 18, 21, 22, 32, 45, 59, 61, 64, 66, 68\\
\hline
$7$  &  30, 37\\
\hline
$8$  &  40, 62\\
\hline
$9$  &  28, 41, 51, 71, 75\\
\hline
$10$  &  74\\
\hline
$12$  &  20, 60\\
\hline
$15$  &  39\\
\hline
$18$  &  73\\
\hline
\hline
\end{tabular}
\begin{flushleft}
$d$: the rank of the abelian group formed by the set of band structures.
\end{flushleft}
\end{table}
\end{center}

%% file: tab_RG_SpinfulNoTRd.tex
\begin{center}
\begin{table}[h]
\caption{
\bf Characterization of band structures for quasi-one-dimensional systems with significant spin-orbit coupling and  no time-reversal symmetry.}
\begin{tabular}{c|c} \hline \hline
$d$ & Rod groups \\
\hline
$1$  &  1, 5, 9, 13, 15, 16, 17, 18, 24, 26, 32, 35, 43, 44, 54, 58, 64, 66\\
\hline
$2$  &  4, 8, 25, 30, 34, 36, 50, 55, 57, 70\\
\hline
$3$  &  2, 3, 7, 10, 12, 14, 19, 20, 21, 22, 31, 33, 37, 38, 41, 42, 47, 48, 49, 56, 62, 63, 67, 68, 69\\
\hline
$4$  &  23, 46, 65, 71\\
\hline
$6$  &  6, 11, 27, 29, 39, 40, 52, 53, 72, 75\\
\hline
$9$  &  45, 51, 59, 61, 73, 74\\
\hline
$12$  &  28\\
\hline
$18$  &  60\\
\hline
\hline
\end{tabular}
\begin{flushleft}
$d$: the rank of the abelian group formed by the set of band structures.
\end{flushleft}
\end{table}
\end{center}

%% file: tab_RG_SpinlessNoTRd.tex
\begin{center}
\begin{table}[h]
\caption{
\bf Characterization of band structures for quasi-one-dimensional systems of spinless fermions without time-reversal symmetry.}
\begin{tabular}{c|c} \hline \hline
$d$ & Rod groups \\
\hline
$1$  &  1, 5, 9, 24, 26, 43, 44, 54, 58\\
\hline
$2$  &  4, 8, 16, 17, 25, 50, 55, 57\\
\hline
$3$  &  2, 3, 7, 10, 12, 14, 19, 31, 33, 35, 36, 42, 47, 48, 49, 56, 63, 67, 70\\
\hline
$4$  &  15, 23, 46, 65, 69\\
\hline
$5$  &  34\\
\hline
$6$  &  6, 11, 13, 18, 21, 22, 27, 29, 32, 38, 52, 53, 64, 66, 68, 72\\
\hline
$7$  &  30, 37\\
\hline
$8$  &  62\\
\hline
$9$  &  40, 41, 45, 51, 59, 61, 71, 75\\
\hline
$12$  &  20, 28, 74\\
\hline
$15$  &  39\\
\hline
$18$  &  60, 73\\
\hline
\hline
\end{tabular}
\begin{flushleft}
$d$: the rank of the abelian group formed by the set of band structures.
\end{flushleft}
\end{table}
\end{center}

%% file: tab_LG_SpinfulTRX.tex
\begin{center}
\begin{table}[h]
\caption{
\bf Symmetry-based indicators of band topology for quasi-two-dimensional systems with significant spin-orbit coupling and time-reversal symmetry.}
\begin{tabular}{c|c} \hline \hline
$X_{\rm BS}$ & Layer groups \\
\hline
$\mathbb Z_{2}$  &  2, 6, 7, 14, 15, 16, 17, 18, 37, 38, 39, 40, 41, 42, 43, 44, 45, 46, 47, 48, 52, 62,\\
                 &  64, 66, 71, 72\\
\hline
$\mathbb Z_{3}$  &  74, 78, 79\\
\hline
$\mathbb Z_{4}$  &  51, 61, 63\\
\hline
$\mathbb Z_{6}$  &  75, 80\\
\hline
\hline
\end{tabular}
\begin{flushleft}
$X_{\rm BS}$: the quotient group between the group of band structures and atomic insulators.
\end{flushleft}
\end{table}
\end{center}

%% file: tab_LG_SpinlessTRX.tex
\begin{center}
\begin{table}[h]
\caption{
\bf Symmetry-based indicators of band topology for quasi-two-dimensional systems with negligible spin-orbit coupling and time-reversal symmetry.}
\begin{tabular}{c|c} \hline \hline
$X_{\rm BS}$ & Layer groups \\
\hline
$\mathbb Z_{2}$  &  2, 3, 7, 49, 50, 52, 66, 73\\
\hline
$\mathbb Z_{2} \times \mathbb Z_{2}$  &  6, 51, 75\\
\hline
\hline
\end{tabular}
\begin{flushleft}
$X_{\rm BS}$: the quotient group between the group of band structures and atomic insulators.
\end{flushleft}
\end{table}
\end{center}

%% file: tab_LG_SpinfulNoTRX.tex
\begin{center}
\begin{table}[h]
\caption{
\bf Symmetry-based indicators of band topology for quasi-two-dimensional systems with significant spin-orbit coupling and no time-reversal symmetry.}
\begin{tabular}{c|c} \hline \hline
$X_{\rm BS}$ & Layer groups \\
\hline
$\mathbb Z_{2}$  &  2, 3, 7, 37, 40, 44, 47\\
\hline
$\mathbb Z_{3}$  &  65, 78, 79\\
\hline
$\mathbb Z_{4}$  &  49, 50, 52, 61, 63\\
\hline
$\mathbb Z_{6}$  &  66, 73, 80\\
\hline
$\mathbb Z_{2} \times \mathbb Z_{2}$  &  6\\
\hline
$\mathbb Z_{3} \times \mathbb Z_{3}$  &  74\\
\hline
$\mathbb Z_{4} \times \mathbb Z_{4}$  &  51\\
\hline
$\mathbb Z_{6} \times \mathbb Z_{6}$  &  75\\
\hline
\hline
\end{tabular}
\begin{flushleft}
$X_{\rm BS}$: the quotient group between the group of band structures and atomic insulators.
\end{flushleft}
\end{table}
\end{center}

%% file: tab_LG_SpinlessNoTRX.tex
\begin{center}
\begin{table}[h]
\caption{
\bf Symmetry-based indicators of band topology for quasi-two-dimensional systems of spinless fermions without time-reversal symmetry.}
\begin{tabular}{c|c} \hline \hline
$X_{\rm BS}$ & Layer groups \\
\hline
$\mathbb Z_{2}$  &  2, 3, 7\\
\hline
$\mathbb Z_{3}$  &  65\\
\hline
$\mathbb Z_{4}$  &  49, 50, 52\\
\hline
$\mathbb Z_{6}$  &  66, 73\\
\hline
$\mathbb Z_{2} \times \mathbb Z_{2}$  &  6\\
\hline
$\mathbb Z_{3} \times \mathbb Z_{3}$  &  74\\
\hline
$\mathbb Z_{4} \times \mathbb Z_{4}$  &  51\\
\hline
$\mathbb Z_{6} \times \mathbb Z_{6}$  &  75\\
\hline
\hline
\end{tabular}
\begin{flushleft}
$X_{\rm BS}$: the quotient group between the group of band structures and atomic insulators.
\end{flushleft}
\end{table}
\end{center}